\documentclass[useAMS,usenatbib]{mn2e}
\usepackage{amstext}
\usepackage{amsmath, float, amssymb}
\usepackage{epstopdf}
\usepackage{graphicx}
\usepackage{txfonts}

\def\simlt{\mathrel{\hbox{\rlap{\hbox{\lower4pt\hbox{$\sim$}}}\hbox{$<$}}}}
\def\simgt{\mathrel{\hbox{\rlap{\hbox{\lower4pt\hbox{$\sim$}}}\hbox{$>$}}}}

\newcommand\be{\begin{equation}}
\newcommand\ee{\end{equation}}
\newcommand\ba{\begin{eqnarray}}
\newcommand\ea{\end{eqnarray}}

\newcommand\FF{{\cal F}}
\newcommand\LL{{\cal L}}
\newcommand\HH{{\cal H}}
\newcommand\SH{{\cal S}}

\title[Relativistic MHD Winds from Rotating Neutron Stars]{Relativistic MHD 
Winds from Rotating Neutron Stars}
\author[N.~Bucciantini et al.]
{
N.~Bucciantini$^{1}$\thanks{niccolo, arons, eliot@astron.berkeley.edu},
T.~A.~Thompson$^{1}$\thanks{Hubble Fellow; Current Address: Department of 
Astrophysical Sciences, Peyton Hall, Ivy Lane, Princeton, NJ 08544, USA; thomp@astro.princeton.edu},
J.~Arons$^{1}$\footnotemark[1],
E. Quataert$^{1}$\footnotemark[1], \&
L.~Del Zanna$^{2}$ \\
$^{1}$Astronomy Department, University of California at Berkeley, 
601 Campbell Hall, Berkeley, CA 94720-3411, USA\\
$^{2}$Dipartimento di Astronomia e Scienza dello Spazio, 
Universit\`a di Firenze, L.go E.~Fermi 2, 50125 Firenze, Italy
}
\begin{document}

\date{Submitted to MNRAS 15 October, 2005}

\pagerange{\pageref{firstpage}--\pageref{lastpage}} \pubyear{2005}
\maketitle
\label{firstpage}


\begin{abstract}
We solve for the time-dependent dynamics of axisymmetric, general relativistic
magnetohydrodynamic winds from rotating neutron stars. 
The mass loss rate as a function of latitude is obtained
self-consistently as a solution to the MHD equations, subject to
a finite thermal pressure at the stellar surface.  We consider both
monopole and dipole magnetic field geometries and we explore the
parameter regime extending from low magnetization 
(low-$\sigma$), almost
thermally-driven winds to high magnetization (high-$\sigma_0$),
relativistic Poynting-flux dominated outflows ($\sigma = B^2/4\pi \rho \gamma c^2 
\beta^2\approx\sigma_0/\gamma_\infty, \; \beta =v/c$ with 
$\sigma_0=\omega^2\Phi^2/\dot{M}$, where $\omega$ is 
the rotation rate, $\Phi$ is the open magnetic flux, and $\dot{M}$ is the
 mass flux).  We compute the angular
momentum and rotational energy loss rates as a function of $\sigma_0$
and compare with analytic expectations from the classical theory of
pulsars and magnetized stellar winds. In the case of the monopole, our
high-$\sigma_0$ calculations asymptotically approach the analytic
force-free limit.  If we define the spindown rate in terms of the open
magnetic flux, we similarly reproduce the spindown rate from recent
force-free calculations of the aligned dipole.  However, even for
$\sigma_0$ as high as $\sim20$, we find that the location of the
$Y$-type point ($r_Y$), which specifies the radius of the last closed 
field line in the equatorial plane, is not the radius of the light cylinder
$R_L=c/\omega$ ($R$ = cylindrical radius), as has previously been assumed in 
most estimates and force-free calculations.  Instead, although the Alfv\'en 
radius at intermediate latitudes quickly approaches $R_L$ as  $\sigma_0$ exceeds 
unity, $r_Y$ remains significantly less than $R_L$. In addition, $r_Y$ is a weak 
function of $\sigma_0$, suggesting that  high magnetizations may be required to
quantitatively approach the force-free magnetospheric structure, with $r_Y=R_L$. 
Because $r_Y<R_L$, our calculated 
spindown rates thus exceed the classic ``vacuum dipole'' rate: equivalently,
for a given spin-down rate, the corresponding dipole field is smaller than traditionally
inferred. In addition, 
our results suggest a braking index generically less than 3.  We discuss the 
implications of our results for models of rotation-powered pulsars and magnetars, 
both in their observed states and in their hypothesized rapidly rotating initial 
states.
\end{abstract}

\begin{keywords}
magnetohydrodynamics (MHD) --- stars: winds, outflows --- relativity --- 
stars: neutron --- methods: numerical
\end{keywords}


\section{Introduction}
\label{section:introduction}

Magnetically dominated winds from stars and accretion disks are central
to the angular momentum evolution of these objects.  Because they can 
efficiently extract rotational energy --- transforming stored gravitational
binding energy into asymptotic wind kinetic energy --- magnetic outflows are 
ubiquitous in powering a wide variety of astrophysical systems.
\citet{sch62} first introduced the key idea that a magnetic field
anchored in a rotating star with a wind can enforce plasma co-rotation
out to radii large compared to the stellar radius, thus greatly
increasing the angular momentum lost per unit mass.  

Schatzman's ideas were
formalized in the steady non-relativistic flow model of \citet{weber67}, who
assumed a pure monopole field geometry, and then by Mestel (1968ab) 
who assumed a more realistic dipole magnetic field configuration.  For
strong dipole fields, a closed zone forms at low latitudes and the
mass loss is concentrated at high latitudes in regions where the
field lines can be opened. If the extent and shape of the open field line 
region is known, then the physics of the magnetic wind is similar to that in the 
monopole geometry: the flow emerges along the open flux tubes and its
character is determined by passing through the slow magnetosonic (SM),
Alfv\'en (AL), and fast magnetosonic (FM) surfaces. 
If the thermal sound speed is small, the flow is driven primarily  by
magneto-centrifugal forces and the asymptotic outflow velocity is 
approximately $v_A (R_A)\approx R_A\omega$,
where $R$ refers to the cylindrical radius, $\omega$ is the stellar
angular velocity, and $R_A$ is the AL radius.  Because $R_A$ is
the point at which $B^2(R_A)/4\pi = \rho v_A^2(R_A)$, 
the outflow kinetic energy is roughly equal to the magnetic energy.
This remains true in the asymptotic wind.

The theory of magnetic winds has been
greatly extended and applied in its non-relativistic form to a wide
variety of problems in stellar astrophysics. \citet{gosling96} and
\citet{asch01} provide reasonably modern reviews in the
solar context, where many of the recent developments in
non-relativistic wind modeling have occurred. Multi-dimensional
models have been motivated in part by the fact that 
toroidal magnetic fields can collimate the flow along the
rotation axis via hoop stress into jets, a topic of much interest in the
modeling of the bipolar outflows observed from protostars (e.g., \citealt{smi98}).

The discovery of rotation powered pulsars motivated the extension of
these ideas to relativistic flows and relativistically strong magnetic
fields ($B^2 /4\pi \gg \rho c^2$ at the source), in order to account
for the observed spin down of pulsars.  \citet{michel69} and
\citet{gj70} made the first extensions of the Weber \&
Davis model to relativistic outflows, recovering the same sequence of
critical points in the flow.  As reviewed in \S \ref{sec:physmod}, these 
studies yielded estimates of the spindown torque of
\begin{equation}
T_\parallel =  k \frac{\mu^2 \omega^3}{c^3}, \; k\sim 1.
\label{eq:reltorque}
\end{equation}
The dipole moment of the star, $\mu$, was assumed to be
related to the monopole moment $m$ appearing in the MHD theory via 
$m = \mu /R_L$, a relation equivalent to assuming that the AL
radius $R_A$ was equal to the radius of the light cylinder 
$R_L = c/\omega$.
Since these monopole models only treated the flow in the
rotational equator, the numerical coefficient $k \sim 1$ in equation
(\ref{eq:reltorque}) could not be determined accurately.
 
These simplified theoretical models revealed important differences
between relativistic and non-relativistic winds. First, 
instead of reaching approximate energy equipartition between flow
kinetic energy and magnetic energy, the asymptotic flow remains strongly magnetized.  
The asymptotic Lorentz factor is given by $\gamma_\infty \approx \sigma_0^{1/3}$, 
where $\sigma_0 = \Phi^2 \omega^2 / \dot{M} c = B_\phi^2/4 \pi \rho c^2$, 
$\Phi$ is the magnetic flux, and $\dot{M}$ is the mass loss rate. Thus,
$\gamma_\infty\ll\sigma_0$.  For a highly relativistic outflow 
($\sigma_0 \gg 1$) the asymptotic magnetization 
$\sigma_\infty = \sigma_0/\gamma_\infty \approx \sigma_0^{2/3} \gg 1$.
A second
important difference is that in relativistic magnetized flows the
electric force cannot be neglected\footnote{In relativistic MHD the electric 
field arises because of motion of the conducting fluid across the magnetic field.
Equivalently, the fluid moves with the single particle 
${\boldsymbol E} \times {\boldsymbol B}$ drift velocity. Therfore, one can eliminate 
the electric field by puting oneself
in the local fluid rest frame and describing the electromagnetic stress as being
solely due to the comoving magnetic field. Occasionally, debate occurs about which
is the ``correct'' frame in which to describe the forces, since workers used to 
non-relativistic MHD are sometimes uncomfortable with the explicit appearance of the
electric field.  When the fluid dynamics is expressed in covariant form (the electrodynamics is already covariant), as we have done in this paper, the forces are 
well defined and unambiguously described in every reference frame.  The choice is dictated only by convenience in describing
the physics, or in actually carrying out the calculations.}.  
Although it is absent by symmetry
in the Michel and Goldreich \& Julian models, the electric force
almost exactly cancels the focusing hoop stress in multi-dimensional
monopole models, thus undermining these flows' usefulness for the
understanding of relativistic collimated jets.  This is a problem generic to all
relativistic outflows \citep{le01} not focused by some
external medium (e.g., a disk or external channel).

Much effort has gone into relaxing the simplifications of
these early models, and in particular, on understanding what is required for an ideal
MHD flow to have $\gamma_\infty \rightarrow \sigma_0$.
Observations and models of pulsar wind nebulae suggest that at the
termination shock of young rotation-powered pulsars' outflows the
magnetic energy has been almost completely converted into flow kinetic
energy (the ``$\sigma$ problem'') and the flow 4-velocity has reached
$\gamma \approx \sigma_0$ (the ``$\gamma$ problem''), in contradiction with
the predictions of the one dimensional monopole treatments. If the magnetic
surfaces retain an almost monopolar shape the acceleration of the flow
is only logarithmic (\citealt{beg94}).  However various asymptotic 
solutions (\citealt{beg94}, \citealt{hey03} and references therein) and similarity 
solutions (\citealt{vl03}, \citealt{vl04}) of parts of the full axisymmetric flow 
problem suggest that if the poloidal magnetic field is sufficiently
non-monopolar beyond the FM point, the largely radial
current that supports the toroidal magnetic field, might cross field
lines toward the equator, causing a transfer of electromagnetic energy
to flow kinetic energy. However, both  
a numerical solution in axisymmetric non-relativistic MHD (\citealt{sak85}) and
a perturbative calculation of relativistically magnetized MHD flow
from the exact force-free solution (\citealt{beskin98}) yield 
a poloidal magnetic field which hews closely to the exact monopolar form.
Thus, the monopole puzzle of asymptotic winds with magnetic energy that is never 
converted to flow kinetic energy persists, suggesting that ideal MHD expansion
of the wind contradicts the observations.

The monopole geometry for the poloidal field has long been recognized as a 
singular case.  A dipole magnetic field aligned with the 
stellar rotation axis presents the simplest realistic alternative field geometry. 
Because the open field, where outflow occurs, has a nozzle shape that expands faster than 
$r^{2}$ at distances comparable to the radius of the last closed field line ($r_Y$), 
there is a possibility of faster acceleration and larger conversion of magnetic energy into 
kinetic energy than occurs in the monopole geometry. However, complete solutions 
in axisymmetric steady flow have only been possible for the monopole; the
change of topology between closed and open field lines required in the
dipole case has so far escaped solution of the Grad-Shafranov equation
that describes the magnetic structure in MHD flows with inertia and pressure included.

The location of the AL radius in the outflow, and its relation
to the maximum equatorial radius of the last closed field line $r_Y$, where
typically a Y-type neutral point occurs in the magnetic field, is a problem of
equal significance.  As outlined in \S \ref{sec:physmod}, if $r_Y <
R_L$, and if $r_Y/R_L$ is an appropriate function of $\omega$, one might
be able to account for the observed fact that in rotation powered pulsars 
$T_\parallel \propto \omega^n$, with $n < 3$.  The strong inferred magnetization of 
pulsars has always suggested that the outflow structure should be treated as a
problem in force-free relativistic MHD, with inertial and pressure forces
completely neglected \citep{gj69}.  Solution of the
force-free Grad-Shafranov equation for the case of an aligned dipole
(\citealt{sch73}, \citealt{michel73}) resisted solution 
until recently (\citealt{cont99} and \citealt{gruz05} assumed strictly force-free
conditions everywhere, whereas \citealt{good04} included pressure effects near the Y-point).
 Gruzinov found that the torque is indeed given by equation
(\ref{eq:reltorque}), with $k=1\pm 0.1$, while Contopoulos et al.~found $k = 1.85$.
All of these authors {\it assumed} $r_Y = R_L$, as is physically plausible.  

However, subsequent work by \citet{timo05} has found that force-free MHD allows a family of
solutions with $r_Y$ of the last field line within $R_L$, contrary to naive expectations. 
He found that $k \approx 0.47 (R_L /r_Y)^2$, which, when $r_Y = R_L$, is
close to Michel's force-free monopole result.    
Importantly, these solutions
indicate that beyond a few Light Cylinder radii, the poloidal magnetic
field assumes a monopolar form, which suggests that acceleration in
the asymptotic wind never reaches equipartition energies (the $\sigma$
problem).  However, since inertial forces are important beyond
the FM surface ($ r \approx \sigma_0^{1/3} R_L$) the force-free approximation breaks down
(\citealt{beskin98}; \citealt{arons04}) and no strong conclusions may be drawn.

The ambiguities of the steady force-free models, and the difficulty of
solving the magnetospheric structure in MHD with inertia and
pressure included, suggest that an evolutionary approach to the
problem would be useful.  A time-dependent solution can then be sought
for specified boundary conditions at the stellar
surface (such as pressure or injection velocity).  The
system can then relax to a self-consistent steady state (if one exists)
and this approach allows one to find the last closed flux surface unambiguously, 
for specified injection conditions.  Furthermore, such a 
treatment allows for the possibility of intrinsically time-dependence
(including either limit-cycle or chaotic behavior) of the magnetosphere.
The time dependence of individual radio pulses from rotation-powered
pulsars, showing systematic drifting through the pulse window for many
objects near the pulsar death line, and chaotic rotation phase over
wider regions of the $P-\dot{P}$ diagram (\citealt{rankin86}, \citealt{desh99})
have timescales consistent with global magnetospheric
variability causing changes in the polar current system underlying
pulsar emission (\citealt{arons81}; \citealt{wright01}).  
Indeed, the torque noise exhibited
by many pulsars \citep{cordes80} may also owe its origin to
instabilities of the magnetospheric current system (\citealt{arons81};
\citealt{cheng87a, cheng87b}).

In this paper we carry out time-dependent RMHD modeling of both highly
magnetized Poynting-flux dominated winds ($\sigma_0\gg1$) in which
$R_A\approx R_L$, as well as magnetized neutron star winds in which $R_A$
is significantly less than $R_L$ ($\sigma_0\ll1$).  

Large $\sigma_0$ MHD models draw their motivation from studies of rotation-powered 
pulsars and of magnetars (Soft Gamma Ray Repeaters and
Anomalous X-ray Pulsars).  More broadly, analogous problems appear in
Poynting-flux dominated models of jets from black holes and neutron stars.
Low-$\sigma_0$ (but still magnetized) neutron star winds are of interest
primarily in understanding the physics of very young neutron stars. In
the seconds after collapse and explosion, the neutron star is hot
(surface temperatures of $\sim$$2-5$ MeV) and extended.  This
proto-neutron star cools and contracts on its Kelvin-Helmholtz
timescale ($\tau_{\rm KH}\sim10-100$\,s), radiating its gravitational
binding energy ($\sim$$10^{53}$ ergs) in neutrinos of all species
(\citealt{burrows86}, \citealt{pons99}).  The cooling epoch is
accompanied by a thermal wind, driven by neutrino energy deposition
(primarily $\nu_en\rightarrow pe^-$ and $\bar{\nu}_ep\rightarrow
ne^+$), which emerges into the post-supernova-shock ejecta  (e.g. \citealt{ducan86}, 
\citealt{qian96}, \citealt{thom01}).

A second or two after birth, the thermal pressure at the edge of the
protoneutron star surface, where the exponential atmosphere joins the
wind ($r_{\nu}$), is of order $\sim10^{28}$ ergs cm$^{-3}$ and decreases sharply
as the neutrino luminosity ($L_\nu$) decreases on a timescale
$\tau_{\rm KH}$, as the star cools and deleptonizes.  The thermal
pressure at the stellar surface is set by $L_\nu$.  If the
protoneutron star has a surface magnetic field of strength $B_0$, then
at some point during the cooling epoch the magnetic energy density
will exceed the thermal pressure.  For fixed $B_0$, the wind region
becomes increasingly magnetically dominated as $L_\nu$ decreases.  For
larger $B_0$ the magnetic field dominates at earlier times. For
magnetar-strength surface fields ($B_0\sim10^{14}-10^{15}$ G) the
magnetic field dominates the wind dynamics from just a few seconds
after the supernova \citep{thom03}.

Magnetars are thought to be born with millisecond rotation periods
(\citealt{duc92}, \citealt{thom93}), in which case
the combination of rapid rotation and strong magnetic fields makes the
proto-neutron star wind magneto-centrifugally driven.  Because the
rotational energy of a millisecond magnetar is very large
($\sim$$2\times10^{52}$ ergs) relative to the energy of the supernova
explosion ($\sim$$10^{51}$ ergs) and because the spindown timescale
$\tau_J\sim\omega/\dot{\omega}\sim(2/5)(M/\dot{M})(r_{NS}/r_A)^2$ 
(where $r_{NS}$ is the neutron star radius) can be
short for large $r_A$, these proto-magnetar winds have been considered
as a mechanism for producing hyper-energetic supernovae 
\citep{thom04}.  Because the wind becomes increasingly
magnetically-dominated and the flow eventually becomes Poynting-flux
dominated as the neutrino luminosity abates, the outflow must become
relativistic.  For this reason, proto-magnetar winds have also been
considered as a possible central engine for long-duration gamma ray
bursts (GRBs) (\citealt{usov92}, \citealt{thom94}, \citealt{wheeler00}, \citealt{thom04}).
They may also be the source of ultra-high energy cosmic
rays  (\citealt{blasi00}, \citealt{arons03}).  The time dependent RMHD models
of neutron star winds calculated in this paper provide a significantly
improved understanding of proto-magnetar winds, and their possible
role in hyper-energetic supernovae and GRBs.


\subsection{This Paper}

With these goals in mind, in \S\ref{sec:physmod} we first outline some
order-of-magnitude estimates for the spindown of neutron stars in both
the low- and high-$\sigma_0$ limits.  In \S\ref{section:math} \&
\S\ref{section:method}, we describe the details of our numerical
model.  In \S\ref{section:results} we present our numerical results
for the 1D monopole (\S\ref{section:1dmonopole}), the 2D monopole
(\S\ref{section:2dmonopole}), and the aligned dipole
(\S\ref{sec:dipole}). In each case, we present a set of models for
both low- and high-$\sigma_0$ winds.  
Finally, in \S\ref{section:discussion} we present a discussion
of our results and speculate on their implications for the spindown of
rotation-powered pulsars and very young, rapidly rotating magnetars.

 
\section{Physical Model}
\label{sec:physmod}
As a guide to the numerical models, we describe here several simple
order of magnitude estimates of the properties of both high- and 
low-$\sigma_0$ winds from neutron stars.
 

\subsection{Poynting Flux Dominated Spindown}
\label{sec:poynting}

\begin{figure}
\hspace*{-2cm}\includegraphics[width=9cm,angle=-90,clip=true]{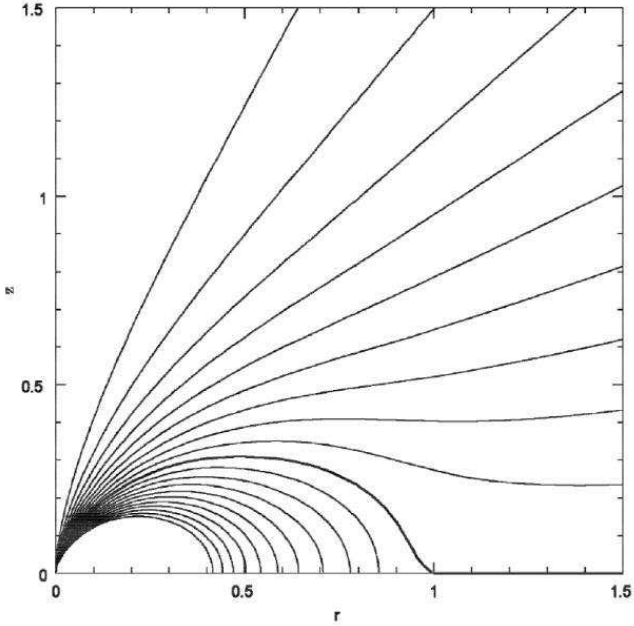}
\caption{Force-free poloidal magnetic field lines from 
a magnetized star with dipole axis aligned with
the rotation axis.  The distances are scaled in units 
of the radius of the Light Cylinder,
$R_{LC}$. From \citet{gruz05}.}
\label{fig:poloidal}
\end{figure}

Consider a star with a magnetic dipole moment ${\boldmath \mu}$. For simplicity, assume
${\boldmath \mu} \parallel {\boldmath \omega}$.  Suppose there is an outflow of plasma
along open field lines which connect to the star in a polar cap, with the magnetic flux 
of the open field lines being $\Phi_o$. The expected poloidal magnetic structure is shown 
in Figure \ref{fig:poloidal}. 
In the closed zone, plasma co-rotates, and the toroidal currents, 
composed of co-rotating charge density and pressure and inertial drifts 
across the magnetic field, cause the distortions from the 
``vacuum dipole'' field, which are of importance at radii comparable 
to $r_Y$.  Assume the AL radius $R_A$ is comparable to $r_Y$.  
The Poynting flux is  ${\boldmath S} = (c/4\pi) {\boldmath E} \times {\boldmath B}$,
whose radial component is, with the poloidal electric field  
${\boldmath E} = - ({\boldmath \omega} \times {\boldmath r}) \times {\boldmath B}_{p}$,
$S_r = -(\omega r \sin \theta /c ) B_p$.  With $ B_{p} (r_A) \approx B_{dipole} (r_A)$ 
and $B_\phi (r_A \sin \theta) \approx - B_p (R_A)$, $S_r (R_A) \sim (\omega R_A /4\pi ) 
\mu^2 /R_A^6$, where subscripts $p$ and $\phi$ denote the poloidal and toroidal 
components, respectively. 
Then the EM spindown torque is approximately 
\begin{equation}
T_\parallel \sim \frac{4\pi r_A^2}{\omega}(2/3)  S_r(R_A) \sim \frac{2}{3}\frac{ \mu^2}{r_A^3} 
        =\frac{2}{3} \frac{\omega^3 \mu^2}{c^3} \left (\frac{R_L}{R_A} \right)^3;
\label{eq:EMtorque}
\end{equation}
the evaluation of the constant $k$ in equation (\ref{eq:reltorque}) to be equal to 2/3 is 
the exact result for the EM torque on the force-free monopole
\citep{michel73}, which has no closed zone or $Y$ point.  Since the open magnetic flux
is $\Phi_o \approx (\mu \omega /c) (R_L/r_Y)$, the torque can also be written as
$T_\parallel \approx (2/3) (\Phi_o^2\omega /c) (r_y /R_L)^2.$

If mass loading (inertial forces) and pressure are negligible, the
long standing expectation has been $r_Y = R_A = R_L = c/\omega$
\citep{gj69}, and MHD spindown torques then should have
braking index $n = \dot{\omega}\ddot{\omega}/\dot{\omega}^2 = 3$,
if $\mu$ and the stellar moment of inertia are constant.  Two
numerical solutions for steady flow from the force-free aligned dipole
rotator have found $k=1.85$ \citep{cont99} and, at
higher resolution, $k=1\pm0.1$ \citep{gruz05}, rather than 2/3.  
The equality $r_Y = R_L = R_A$ was assumed in both of these calculations.

A long standing empirical puzzle has been that in the four
observations of braking indices not requiring major corrections for
glitches in the timing, the braking index lies between 2.5 and 2.9
 (\citealt{lyne93}, \citealt{kaspi94}, \citealt{deeter99}, \citealt{camilo00},
\citealt{livi05}). This reduction of the braking index
for fixed $\mu$ and $\omega$, in comparison to our simple estimate, is 
tantamount to $r_Y < R_L$.  That is, the closed zone ends within the
Light Cylinder\footnote{An alternate possibility is an increasing
magnetic moment \citep{bland88}. Still another option is evolution of the
angle between ${\boldsymbol \omega}$ and ${\boldsymbol \mu}$, in the still
unassessed dependence of the electromagnetic torque on the oblique rotator, in 
either force-free or RMHD models.}, {\it and}, as the star ages ($\omega$
decreases), $r_Y / R_A$ also decreases --- the magnetosphere becomes
more open with decreasing spin down power.  One way to state this is to
simply assert that $R_A = r_Y < R_L$ and 
that $r_Y / R_L\propto \omega^\alpha$ (e.g., $r_Y/R_L = (\omega r_{NS}
/c)^\alpha$, \citealt{arons83}).  This implies a change in the polar cap size from $r_{NS}
(r_{NS}/R_L)^{1/2}$ to the larger value (larger $\Phi_o$) of 
$r_{NS}(r_{NS}/R_L)^{(1-\alpha)/2}$. Using this expression, the braking index data
require $1/6 \geq \alpha \geq 1/30$, with the largest value for the
Crab pulsar and the smallest for the 407 ms pulsar J1119-6157.

Assuming $\alpha > 0$ is equivalent to the last closed field line of the
dipole having equatorial radius $r_Y$ less than $R_A \approx R_L$.  According to our 
estimate (\ref{eq:EMtorque}), the electromagnetic torque depends on the field
strength at $R_A$, which is noticeably larger than that estimated by using a pure dipole
filed, since the poloidal field becomes progressively more monopolar for $r > r_Y$.
Thus one obtains a better estimate of the torque by using a simplified model of the 
poloidal magnetic field
which has the correct asymptotic form shown by the force-free aligned rotator 
models - dipolar
at $r \ll r_Y$ and monopolar at $r \gg r_Y$.  Our RMHD results have the same
asymptotic behavior.  Thus, with $B_p = (\mu /r^3) + \kappa (\mu /r_Y r^2)$, the same
order of magnitude argument that led to (\ref{eq:EMtorque}) yields
\begin{eqnarray}
T_\parallel & = & k  \frac{\omega^3 \mu^2}{c^3} \left (\frac{R_L}{R_A} \right)^3 ( 1 + f)^2, \\
          f & \equiv & \kappa \frac{R_A}{r_Y} .
\end{eqnarray}
Assuming the magnetic moment and the stellar moment of inertia (and
$i = \angle ({\boldsymbol \omega}, \boldsymbol{\mu})$ are constant, and that
$R_A = R_L$, a correspondence assumed in most force free models 
and also found in  the RMHD results we report below, one readily finds
\begin{equation}
n = \frac{\Omega \ddot{\omega}}{\dot{\omega}^2 } =
        3 + 2 \frac{\partial \ln (1 + f)}{\partial \ln \omega}.
\label{eq:brake_ind}
\end{equation}
Thus a braking index less than 3 requires $R_A /r_Y$ to depend on
$\omega$ (more generally, to  depend on time.) If the time
dependence of $r_Y/R_A$ enters solely through dependence on $\omega$,
$n < 3$ requires   $R_A /r_Y = R_L/r_Y$ to
increase as the star spins down - the magnetosphere 
becomes more open with time, and the magnetic field at the
light cylinder to remain larger, than is expected in the traditional model.
Such behavior requires transformation of closed field lines to open,
which can occur if magnetic dissipation at and near the Y-point allows reconnection 
to enable this transformation. 

Our RMHD numerical results presented in \S\ref{sec:dipole} show that although
$R_A \rightarrow R_L$ as soon as $\sigma_0$ exceeds unity, 
$r_Y / R_L$ remains substantially less than unity for $\sigma_0$ 
as large as $\sim$20.  We also find that, for $\sigma_0$ of order a few,
 the ratio $r_Y / R_L$ decreases as $\omega$ decreases - the braking index 
in our models is less than 3.  
This suggests that seemingly small inertial and pressure forces
can have a large effect on the magnetospheric structure and, in turn, 
the magnitude of the spindown torque and the braking index.

The work of \citet{mestel87} may provide an explanation for our numerical 
results.  In their 
isothermal, non-relativistic analysis of the magnetohydrostatic equilibrium
of the closed zone, they find that $R_Y/r_{NS}$ depends crucially
on the ratios $V^2_{e}/2c_T^2$ and $\omega^2 r_Y^2/V^2_{e}$, 
where $V_{e}$
is the escape velocity from the stellar surface (see their \S2 \&
eq.~8).  For $r_Y$ a few times $r_{NS}$ one finds an approximate implicit
equation for $r_Y$:
\begin{equation}
\left(\frac{r_Y}{r_{NS}}\right)^6=\frac{(B_0^2/8\pi)}{\rho_{NS}c_T^2}
\exp\left[\frac{V^2_e}{2c_T^2}\left(1-\frac{\omega^2r_Y^2}{V_e^2}\right)\right].
\end{equation}
One sees that for fixed isothermal sound speed $c_T$, if $(\omega r_Y/V_{e})$ 
is greater than unity, then it becomes exponentially harder to increase
$R_Y/r_{NS}$ by increasing $B_0^2/\rho_{NS}$.

However, since electromagnetic stresses alone can lead to $Y$ point formation,
as is clear from the solutions of the force-free Grad-Shafranov equation, one should treat these
estimates of $R_Y$ as a function of plasma stress alone with caution, when
applied to the magnetically-dominated regime.  Simplified models along the lines
pioneered by \citet{good04} may be helpful, but at present, the best results are
the simulations themselves (\S\ref{sec:dipole}). We defer a relativistic 
generalization of \citet{mestel87} with a polytropic ($p\propto\rho^\Gamma$) equation of state (more appropriate to our simulations) to a future paper.


\subsection{Spindown by Magnetized Mass-Loaded Winds }
\label{sec:massloaded}

Stresses due to mass loading become 
significant in the thermally driven winds of  newly born 
neutron stars. These stars may have strongly mass-loaded winds which have
$r_A$ large in comparison to $r_{NS}$, but significantly smaller than
$R_L$ (\citealt{thom04}).  Conditions for such winds in the presence of thermally-driven
mass loss generally obtain when the isothermal sound speed $c_T$ is
smaller than, or of order, the Alfv\'en speed, $v_A$.\footnote{For
very high mass loss rates and/or high thermal pressures, $\rho
c_T^2\gg B^2/8\pi$. In this limit, the magnetic field is not important
in accelerating matter off of the stellar surface, the wind is driven
thermally, and spindown is negligible unless the star rotates at a
significant fraction of breakup.}

The torque on the star by the magnetized wind is easily estimated for
a strictly monopolar field geometry ($B_p\propto r^{-2}$).  At $r_A$
we expect the poloidal magnetic energy density to be of order the
radial kinetic energy density, $B_p^2/8\pi\simeq\rho v_r^2/2$, and if
$c_T$ is much smaller than $v_A$, then we further expect that
$v_r(r_A)$ should be of order $v_\phi(r_A)$.  A simple estimate for
$v_\phi(r_A)$, which again requires that $v_A\gg c_T$, is $v_\phi\sim
r_A\omega$.  We combine the above ingredients to obtain an expression
for the Alfv\'en point:
\begin{equation}
r_A=B_0^{2/3}r_{NS}^{4/3}(\dot{M}\omega)^{-1/3},
\end{equation}
where we have assumed that $\dot{M}=4\pi r^2\rho v_r$ and $B_0$ is the 
magnetic field strength at the stellar surface.  The time evolution of the
star's spin period is then governed by the equation
\begin{equation}
\dot{J}=-\dot{M}r_A^2\omega=-B_0^{4/3}r_{NS}^{8/3}(\dot{M}\omega)^{1/3}.
\end{equation}
The rotational energy loss rate is 
$\dot{E}=I\omega\dot{\omega}\propto\dot{M}^{1/3}$.
In this limit ($v_A\gg c_T$) the sonic point is approximately
\begin{equation}
r_{s}=\left(\frac{GM}{\omega^2}\right)^{1/3}
\hspace{-.3cm}=\left(\frac{r_{Sch}R_L^2}{2}\right)^{1/3}
\hspace{-.3cm}\simeq17\,M_{1.4}^{1/3}P_1^{2/3}\,\,{\rm km},
\label{rsonic}
\end{equation}
where we have scaled the spin period $P$ to 1 millisecond and
$r_{Sch}$ is the Schwarzschild radius.  In the $c_T\ll v_A$ limit, the
radial velocity reaches its asymptotic value of $v_\infty\approx (3/2)
v_A$ at the FM  point (e.g.~\citealt{belcher76}).

In the regime we are interested in this paper we always have a supersonic
outflows at the AL surface. In this case, according for the standar papametrization
of MHD winds (\citealt{sak85}, \citealt{daigne02}), all solution should be 
considered centrifugally or marginally centrifugally driven. 

A similar set of estimates for a dipole magnetic field is considerably
more complicated, particularly since, as in \S\ref{sec:poynting}, the
position of the $Y$ point is not known {\it a priori}.  In addition,
since the areal function along open flux tubes adjacent to the closed
zone deviates significantly from radial, a full numerical solution is
required to address spindown in this context (\S\ref{sec:dipole}). 
However, as a rough guide in understanding the expected differences between 
monopole and dipole spindown it is sufficient to imagine the scalings for 
the dipole as essentially those for the monopole, but with the surface 
magnetic field strength normalized to just the open magnetic flux. 


\subsection{Parameterization \label{sec:param}}

These estimates and those of \S \ref{sec:poynting} reveal  the parameters which 
specify the physical regimes of relevance to our models of rotating magnetospheres.  
Mass loss is thermally and centrifugally driven in these models, depending  
upon the ratio of pressure and centrifugal forces to the
gravitational force, parameterized at the injection surface ($r_{in}$)
 by $(c_T^2/V_{e}^2)_{r=r_{in}}$ and by
$(\Omega r_{in}/V_{e})^2_{r=r_{in}}$.  All of the models considered in this
paper have the first ofthese parameters between 0.01 and 0.1, while the second
is between 0.05 and 0.3. The values adopted 
can be derived from the paramenter shown in Tables 1 (\S\ref{sec:init}).  
For all of the 2D monopole and dipole
models, magnetic pressure dominates gas pressure, as expressed by 
$B^2/4\pi c_s^2 > 1$.  This is true for most of the 1D monopole models also.
Again, these parameters are listed in Table 1. In all cases, the thermal energy
density is smaller than the rest mass density ($p < \rho c^2$). Thus, pressure
forces do not lead to relativistic motion.  The values of the ratios between the 
characteristic speeds at the base of the wind, for all our simulations, are provided
in Appendix A. 

The distinction between pressure deriven and centrifugally driven wind, can be 
also done based on the conditions at the AL surface. In \citet{daigne02} 
(following \citet{sak85}) the distincion is based on the value of the ratio 
$\Gamma p/\rho(\Omega r)^2$ at the AL surfaces ($\Gamma$ is the adiabatic 
coefficient). In all our cases such ratio is less that 0.1.

The most significant parameter is Michel's magnetization parameter,
$\sigma_0$, defined just after expression (\ref{eq:reltorque}).  When the
magnetic energy density exceeds the rest mass density ($\sigma_0 > 1$),
magnetic pressure can accelerate the flow to relativistic velocities.  This
parameter is listed for all the models, as our major goal is to span the
regimes from highly mass loaded, pressure driven, nonrelativistic outflow 
($\sigma_0 \ll 1$) to lightly mass loaded, magnetically driven relativistic
outflow ($\sigma_0 \gg 1$) in both monopole and dipole geometry. The values
of $\sigma_0$ for the various models are given in the tables in \S\S
\ref{section:1dmonopole}, \ref{section:2dmonopole} and \ref{sec:dipole}.

We do not consider outflows driven by relativistically high temperature
($p > \rho c^2$), a regime more relevant to fireball models of Gamma Ray
Bursts.


\section{Mathematical Formulation}
\label{section:math}

The laws of mass and momentum-energy conservation, together with Maxwell equations in general 
relativity, are (\citealt{landau71}, \citealt{weinberg72}, \citealt{misner73}, \citealt{anile89}):
\ba
&&\nabla_{\nu}(\rho u^{\nu})=0 \label{eq:grmhd1},\\
&&\nabla_{\nu}(T^{\mu\nu})=0,\\
&&\partial_{\mu}F_{\nu\lambda}+\partial_{\nu}F_{\lambda\mu}+\partial_{\lambda}F_{\mu\nu}=0,\\
&&\nabla_{\mu}(F^{\mu\nu})=-J^{\nu},
\label{eq:grmhd4}
\ea
where $\rho$ is the proper rest mass density, $u^{\nu}$ and $J^{\nu}$ are the four-velocity 
and the four-current density, and $F^{\mu\nu}$ is the Faraday tensor of the electromagnetic field. 
The momentum-energy tensor is given by:
\be
T^{\mu\nu}=\rho h u^\mu u^\nu + g^{\mu\nu}p+F^{\nu\sigma}F^{\mu}_{\sigma}-g^{\mu\nu}F^{\lambda\kappa}
F_{\lambda\kappa}/4,
\label{eq:tmn}
\ee where $g^{\mu\nu}$ is the metric, $p$ is the gas pressure, and we
have chosen a system of units in which $c=1$. In the case of
$\Gamma$-law equation of state for a perfect gas the specific enthalpy
is $h=1+\Gamma/(\Gamma-1)p/\rho$. In order to close the system, the
current density $J^{\nu}$ must be specified in terms of the other
known quantities, through an additional equation, Ohm's law. In the
MHD approximation Ohm's law becomes the condition that the net
electric field in the fluid frame must vanish, which in covariant form
reads: \be F_{\mu\nu}u^\nu =0.
\label{eq:ohm}
\ee 
With this approximation, Eqs.~(\ref{eq:grmhd1})-(\ref{eq:grmhd4}) can be rewritten 
in term of proper density, pressure, four-velocity and magnetic field, reducing to a 
system of 8 equations for 8 variables, plus the solenoidal condition on the
 magnetic field, $\nabla\cdot{\bf B}=0$.

Although we consider winds from neutron stars with centrifugal forces large
enough to affect the mass loss, we only consider rotation rates slow enough
to allow us to neglect rotational modifications of the metric. Therefore,
employ the Schwarzschild instead of the Kerr metric. The use of a
diagonal metric allows one to simplify the equations and to implement
them easily in any code for special relativistic MHD, as shown by
\citet{koide99}. 

In Boyer-Lindquist coordinates ($t$,$r$,$\theta$,$\phi$) the diagonal elements of 
the metric are:
\ba
-g^{00}=(g^{11})^{-1}=(1-2GM/r)=\alpha^2;\\
g^{22}=r^2;\;\;\; g^{33}=r^2\sin{\theta}^2;\nonumber
\label{eq:shm}
\ea where $M$ is the mass of the central object (1.44 $M_\odot$).  

In axisymmetry and steady state, equations
~(\ref{eq:grmhd1})-(\ref{eq:grmhd4}) admit 6 integrals of motion along
stream lines (flux tubes) (\citealt{cam86a}, \citealt{cam86b}, \citealt{cam87}, 
\citealt{daigne02}): 
\ba 
&&\FF=\alpha\rho\gamma v_p A\label{eq:integi},\\
&&\Phi=B_pA,\\ 
&&\Omega=\alpha(v_\phi-B_\phi/B_p v_p)/R,\label{eq:rot}\\
&&\LL=R(h\gamma v_\phi-\alpha\Phi B_\phi/\FF),\\ 
&&\HH=\alpha(h\gamma -\Omega\Phi RB_{\phi}/\FF),\label{eq:by}\\ 
&&\SH=p/\rho^\Gamma,
\label{eq:integf}
\ea where subscript $p$ and $\phi$ indicate poloidal and azimuthal
components, respectively, $\gamma$ is the Lorentz factor, $A$ is the
area of the flux tube ($A=r^2$ in the case of radial outflow), and
$R=r\sin{\theta}$ the cylindrical radius. Again note that we have
chosen units in which $c=1$.

If the value of the six integrals is known on a stream line, then
equations ~(\ref{eq:integi})-(\ref{eq:integf}) can be solved for the
value of primitive quantities like density, velocity, and pressure. In
general, quantities like $\Omega, \SH$, and $\Phi$ are assumed to be
known and given by the physical conditions at the surface of the
central object, while the values of the remaining integrals are
derived by requiring the solution to pass smoothly the
slow-magnetosonic (SM), Alfv\'{e}nic (AL) and fast-magnetosonic (FM)
points. In 2D, one also requires an additional equation for the area
of the flux tube, and this is provided by requiring equilibrium across
streamlines.


\section{Numerical method}
\label{section:method}

Equations (\ref{eq:grmhd1})-(\ref{eq:grmhd4}) are solved using the
shock-capturing code for relativistic MHD developed by \citet{ldz03}.
 The code has been modified to solve the equations in
the Schwarzschild metric following the recipes by 
\citet{koide99}. The scheme is particularly simple and efficient,
since solvers based on characteristic waves are avoided in favor of
central-type component-wise techniques (HLL solver based only on the
FM speed). In the axi\-symmetric 2D approximation the
equation for the evolution of $B_\phi$ can be written in conservative
form and only one component of the vector potential, $A_\phi$, is used
in the evolution of the poloidal magnetic field. Moreover we replace
the energy conservation equation with the constant entropy condition,
$\SH=const$. Of course this condition cannot be satisfied during the
evolution when shocks form in the flow. However, if the wind evolves
toward steady state, shocks propagate outside the computational
domain, and the condition $\SH=const$ can be satisfied. There are
various numerical reasons for our choice of $S = const$. Most
importantly, enforcing constant entropy significantly enhances code
stability. For example, it is known that in strongly magnetized flows
or in the supersonic regime, in deriving the thermal pressure from the
conserved quantities, small errors can lead to unphysical states. In
non-relativistic MHD these unphysical states correspond to solutions
with negative pressure. A common fix is to set a minimum pressure
``floor'' that allows the computation to proceed. However, in
relativistic MHD (RMHD) it is possible that no state can be found (not
even one with negative pressure) and the computation stops. This
usually happens for high Lorentz factor $\gamma\sim10-100$, depending
on the grid-flow geometry, or in the case of high magnetization,
$B^2/(\rho h) \simgt 100$. When the magnetization at the Light
Cylinder is high, close to the central object it will be above this
stability threshold, which prevents us from using the full equation of
energy. The use of a constant entropy condition also greatly
simplifies the derivation of primitive variables from the conserved
quantities, thus increasing the efficiency of the code. 

Lastly, by specifying the value of the entropy we remove entropy waves
from the system. Entropy waves travel at the advection speed (and are
dissipated in an advection time). If the sound speed is much smaller
than $c$, the advection speed $v_r$, close to the surface of the
central object, is much smaller than the speed of light and the
advection time can be extremely long. By removing entropy waves,
perturbations are dissipated at the SM speed, which in our regime is
of order of $0.1\,c$.

Simulations were performed on a logarithmic spherical grid with 200
points per decade in the radial direction, and a uniform resolution on
the $\theta$ direction with 100 grid points between the pole and the
equator (CFL factor equal to 0.4). Higher resolutions were used in a few cases, to test
convergence and accuracy. In order to model the heating and cooling
processes using an ideal gas equation of state we adopt an adiabatic
coefficient $\Gamma=1.1$ (almost isothermal wind), which is reasonably
representative of the wind solution by, e.g., \citet{qian96}. It can be shown
that in order to have a transonic outflow the thermal pressure cannot
be too high (above a critical value depending on $\Gamma$ the sonic
point moves inside the surface of the star) nor too low (the
Bernoulli integral $\HH$ must be positive), as shown by \citet{koide99}
 in the hydrodynamical case. The available parameter
space increases for smaller $\Gamma$. The value we choose allows us to
investigate easily cases with $p/\rho \sim 0.05$, especially in the dipole case
where the strong flux divergence at the base is more important.


\subsection{Initial and boundary conditions, and Model Parameters
  \label{sec:init}}

Simulations in both 1D and 2D were initialized by using a
hydrodynamical 1D radial solution obtained on a much finer grid (the
relativistic extension of the Parker solution 
\citep{parker58}, and projecting it on the initial magnetic field
lines. In the monopole cases, initial poloidal magnetic field lines
are assumed to be strictly radial while for the dipole cases we adopt
the solution for the vacuum dipole in the Schwarzschild metric (i.e. \citealt{muslimov92},
\citealt{wasser83}). Density and pressure were
interpolated from the hydrodynamical solution, $v_r$ was derived by
projecting the radial velocity on the magnetic field lines, and we set
$v_\theta=0$. We also impose co-rotation in the inner region
$v_\phi={\rm min}(\Omega R /\alpha, 0.6c)$, in order to avoid sharp
temporal transients in the vicinity of the inner boundary.

Standard reflection conditions are imposed on the axis, and symmetric conditions are 
imposed on the equator. At the outer radial boundary we apply standard 0-th order 
extrapolation for all the variables. Initial condition are chosen in order to 
guarantee that during the evolution the FM surface is inside the computational 
domain so that no information is propagated back from the outer boundary.
 
Unfortunately, as we will discuss in the following section, in the 2D
case such a constraint can not be satisfied close to the axis, unless
one uses an excessively large computational domain.  
In a few cases, using larger grids that
allow the FM surface to be inside the computational domain, we find
that the results do not change appreciably except along the axis
itself. That is, the global solution at all but the highest latitudes is not
significantly affected by the fact that the FM surface is outside the
computational domain very near the polar axis.

\begin{table*}
\begin{minipage}{17cm}
\caption{Parameters of the numerical models (unit with $c=1$).}
\label{table:1}
\begin{center}
\begin{tabular}{c c c c c c c c c c c}
\hline
1D Mod. & A0 & A & B & C & D & E & F & G & H & I\\
\hline
$c^2_T(r_{in})$ & 0.033 & 0.033 & 0.028 & 0.025 & 0.021 & 0.018 & 0.033 & 0.030 & 0.030 & 0.030\\
$B_r(r_{in})^2/\rho(r_{in})$ & 0.004 & 0.04 & 0.04 & 0.04 & 0.04 & 0.04 & 4 & 8 & 32 & 80 \\
\hline
2D Monop. Mod. & A & B & B1 & C & D & E & & &  \\
\hline
$c^2_T(r_{in})$ & 0.033 & 0.033 & 0.033 & 0.033 & 0.033 & 0.033 &  &  &  \\
$B_r(r_{in})^2/\rho(r_{in})$ & 0.04 & 0.4 & 0.4 & 4 & 40 & 200 &  &  &  \\
\hline
2D Dip. Mod. & A & B & B1 & B2 & C &  &  & & \\
\hline
$c^2_T(r_{in})$ & 0.033 & 0.033 & 0.033 & 0.033 & 0.033 &  &  &  &  \\
$B_r(r_{in},\theta=0)^2/\rho(r_{in})$ & 0.64 & 6.4 & 6.4 & 6.4 & 64 &  &  &  &  \\
\hline
\end{tabular}
\end{center}
\medskip

In all models $\Omega=0.143$ except models $B1$  which have $\Omega=0.214$ and model 
$B2$ which has $\Omega=0.0715$. See Eq.~(\ref{eq:period}) for the corresponding value 
of the rotation period. $c_T^2=\Gamma p/(\rho h)$. Case A0 is a reference case for an 
almost thermally driven wind. In term of the standard wind parametrization 
(\citealt{sak85}, \citealt{daigne02}) all our cases are centrifugally driven: the 
value of $\Gamma p/\rho(\Omega r)^2$ at the AL surfaces, is always less than 0.1 
except in case A0 where it is 0.5. The value of the lapse at the injection radius is
$\alpha(r_{in})=0.79$ correspondig to an escape speed $0.6 c$.
The unit of length corresponds to the radius of the neutron star $r_{NS}$, the unit of time
is $r_{NS}/c$. 

\end{minipage}
\end{table*}

Particular care has to be taken for the inner boundary conditions. As pointed
out in \S\ref{section:introduction}, we are here interested in the transition from 
mass  loaded ($\sigma_0 \ll 1$)
to high-$\sigma_0$ winds, and our injection conditions are tuned to
be as close as possible to the neutrino driven proto-neutron star case. We
chose the inner radius $r_{in}$ to be located at 11 km (1.1 radii of
the neutron star $r_{NS}$), which corresponds to the outer edge of the
exponential atmosphere for a thermally-driven wind (e.g., \citealt{thom01}).
 The modeling of such a steep atmosphere requires very high
resolution in order to avoid numerical diffusion, and the problem
becomes prohibitive in terms of computational time in 2D. At the inner
boundary the flow speed is smaller than the SM speed, implying that
all wave modes can have incoming and outgoing characteristics. This
constrains the number of physical quantities that can be
specified. Density and pressure at the inner radius are set to be
$p/\rho \sim 0.04$, thus fixing the entropy for the overall wind. The
radial velocity is derived using linear extrapolation. We also fixed
the value of $\Omega / \alpha(r_{in})$, typically at
$\Omega r_{in}/\alpha(r_{in}) = 0.2$, corresponding to a millisecond period
(in the neutron star proper frame, see also Eq.~(\ref{eq:period})),
which implies $R_L=6.8 r_{NS}$.

The frozen-in condition Eq.~(\ref{eq:ohm}) requires that the electric
field in the comoving frame at the inner radius vanishes: $v_p
\parallel B_p \rightarrow E_\phi=0$ and $E_p =\Omega
R_{in}B_p/\alpha(r_{in})$. The condition on $E_\phi$ implies that the
radial component of the magnetic field remains constant. We chose for
$B_r(r_{in})$ different values to investigate both cases with low
magnetizations and high magnetization. $B_\theta$ and $B_\phi$ where
extrapolated using 0-th order reconstruction (we found that linear
interpolation can lead to spurious oscillations). The value of the
tangential velocities were derived using the remaining constraint of
the frozen-in condition: $v_\theta= v_r B_\theta/B_r$ and
$v_\phi=\Omega R_{in}/\alpha(r_{in})-v_rB_\phi/B_r$. Given that the 3
components of the velocity are derived independently there is no
guarantee that $v^2 <1$, so care has to be taken to avoid
sharp transients and spurious oscillations in the tangential magnetic
field near the inner boundary.

In Table \ref{table:1} the parameters of our various models are given.
Equations (\ref{eq:integi})-(\ref{eq:integf}) show that the problem
can be parameterized in terms of the ratio $\Phi^2/\FF$ (assuming $B_\phi$
scales as $B_r$) and not on the specific value of density and magnetic
field; more generally the parameter governing the properties of the
system is $\sigma_0=\Phi^2\Omega^2/\FF$ \citep{michel69}. The
bulk of our simulations have been done using a fixed value for
$\Omega$, in order to allow a more straightforward comparison among
the various results; however in a few cases (B1 of the 2D monopole,
and B1, B2 of the 2D dipole) we use a different rotation rate to check
whether the energy and angular momentum loss rates indeed only scale
with $\sigma_0$.  We note that it is computationally more efficient to
increase the magnetic flux than to drop the mass flux in order to
achieve higher $\sigma_0$. Only in the 1D case, where resolution is not
a constraint, are we able to investigate the behavior of the system
for different values of $\SH$.


\section{Results}
\label{section:results}


\subsection{The 1D Monopole}
\label{section:1dmonopole}

As a starting point for our investigation, we consider the simple case
of a relativistic monopolar wind in 1D. This is the relativistic
extension of the classic Weber-Davis solution for a magnetized wind
\citep{weber67}, and represents a simplified model for the flow
in the equatorial plane. The 1D model can also be used both to verify
the accuracy of the code and as a guide in understanding the 2D
simulations (\S\ref{section:2dmonopole}, \ref{sec:dipole}). Here we
assume $v_\theta = B_\theta =0$.  The solenoidal condition on the
poloidal magnetic field reduces to $B_r\propto r^{-2}$.

We calculate a number of wind solutions for different values of the
flux $\Phi$ and the entropy $\SH$. The results are summarized in Table
\ref{table:2}. Case A0 is a reference case for an almost unmagnetized wind, 
and it will be used for comparison with the weakly magnetized regime (A--D).
As stated above, the solution can be parameterized in
terms of $\sigma_0=\Phi^2\Omega^2/\FF$. We have verified that the mass
loss rate $\FF$ depends strongly on the value of the sound speed at
$r_{in}$, and drops rapidly as the pressure approaches the critical
value for $\HH>0$, which also depends on the magnetic field
strength. In contrast, over the range of parameters studied, $\FF$ has
a relatively minor dependence on the value of the magnetic field at
fixed $\SH$: increasing $B_r(r_{in})^2$ by 3 orders of magnitude
corresponds to an increase in $\FF$ by just a factor $\sim1.7$
(compare models A0 and F in Tab.~\ref{table:2}). The reason for this
is that in all cases listed in Table \ref{table:2} the AL point is
larger than the SM point.  Thus, the magneto-centrifugal effect of
increasing the density scale height in the region interior to the SM
point, where the mass loss rate is set, is already maximized. 
For all the cases investigated the value of $\Gamma p/\rho(\Omega r)^2$ at the 
AL point is always less than 0.1 (in case A it is $\simeq 0.1$ while in case 
E it is $\simeq 0.02$). All the solution can thus be considered centrifugally driven.
 We expect and find a sharp drop in $\FF$ as we go to yet smaller
magnetization.  For example, in the purely hydrodynamical case without
magnetic fields, $\FF\approx30$, six times lower than $\FF$ for Case F.

\begin{figure}
\resizebox{\hsize}{!}{\includegraphics[clip=true]{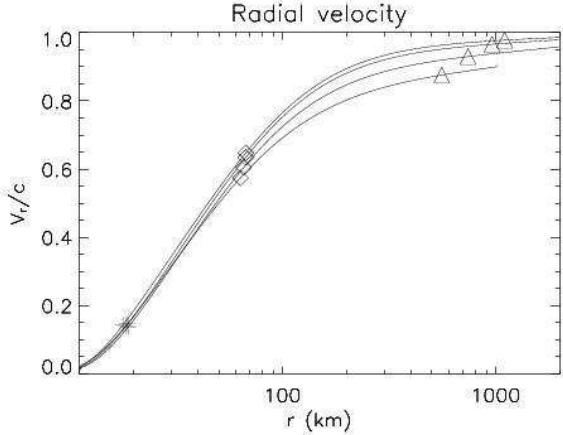}}
\caption{Radial velocity, and position of the SM (plus), AL (diamond) and FM (triangle) points, 
in the 1D monopole case. From bottom to top lines refer to cases F, G, H and I.}
\label{fig:1}
\end{figure}

Note that the value of $\gamma$ reported in Table \ref{table:2} is at
a fixed position, $r=100r_{NS}$, which is generally outside the FM
point (see Fig.~\ref{fig:1}).  Case I is an exception.  For reference,
the Lorentz factor at $\sim300\, r_{NS}$ is 7.8 for this model.

All values in Table \ref{table:2} are given in code units. The
following relations can be used to scale to physical units, in terms
of the value of density and magnetic field at the injection radius
$r_{in}$.  The mass loss rate is 
\be 
\dot{M}=1.9\cdot
10^{-4}\,\FF\left(\frac{\rho(r_{in})}{10^{10} {\rm g\,
cm}^{-3}}\right)\left(\frac{r_{in}}{11\, {\rm km}}\right)^2
M_{\odot}\,{\rm s}^{-1};
\label{eq:c1}
\ee
the rotation period is
\be
P=2.3\cdot 10^{-4} \Omega^{-1}\left(\frac{r_{in}}{11\, {\rm km}}\right)
 {\rm\, s};
\label{eq:period}
\ee
the surface magnetic field strength is
\ba
B_r(r_{in})&=&4.25\cdot 10^{15}\,\sigma_0^{1\over 2}\left(\frac{\dot{M}}{1.9\cdot 10^{-4}\,M_{\odot}\,
{\rm s}^{-1}}\right)^{1\over 2}\times\nonumber\\
&&\hspace*{-1.45cm}\left(\frac{P}{1.6\cdot 10^{-3}{\rm s}}\right)^{1\over 2}\left(\frac{\rho(r_{in})}
{10^{10} {\rm g\, cm}^{-3}}\right)^{-{1\over 2}}\left(\frac{r_{in}}{11\, {\rm km}}\right)^{-1}\,{\rm G};
\ea
the total energy loss rate is
\ba
\dot{E}&=&6.95\cdot\,10^{48}\,\left(\frac{\dot{H}}{\Phi^2\Omega^2}\right)\left(\frac{B_r(R_{in})}
{8.7\cdot 10^{13}\,{\rm G}}\right)^2\times\nonumber\\
&&\left(\frac{R_{in}}{11\, {\rm km}}\right)^2\left(\frac{P}{1.6\cdot 10^{-3}{\rm s}}\right)^{-2}\,
{\rm ergs\, s}^{-1};
\ea
the angular momentum loss rate is
\ba
\dot{L}&=&1.57\cdot 10^{45}\,\left(\frac{\dot{J}}{\Phi^2\Omega}\right)\left(\frac{B_r(r_{in})}
{8.7\cdot 10^{13}\,{\rm G}}\right)^2\left(\frac{r_{in}}{11\, {\rm km}}\right)^3\times\nonumber\\
&&\left(\frac{P}{1.6\cdot 10^{-3}{\rm s}}\right)^{-1}\,{\rm ergs},
\label{eq:cf}
\ea
where $\dot{J}=\LL\FF$ and $\dot{H}=\HH\FF$.

\begin{table}
\caption{Results of the 1D models. }
\label{table:2}
\begin{center}
\begin{tabular}{c c c c c c c c}
\hline
Mod. & $\sigma_0$ & $\FF$ & $\dot{J}/\Phi^2\Omega$ & $\dot{H}/\Phi^2\Omega^2$ & $\gamma(100 r_{NS})$ & $r_A/R_L$\\
\hline
A0 & 0.013 & 101  & 10.6 & 99.0 & 1.12 & 0.30\\
A & 0.084  & 142  & 4.17 & 18.0 & 1.20 & 0.46\\
B & 0.164  & 74   & 3.34 & 9.98 & 1.22 & 0.56\\
C & 0.307  & 39   & 2.71 & 6.06 & 1.32 & 0.65\\
D & 0.634  & 19   & 2.18 & 3.75 & 1.42 & 0.75\\
E & 2.81   & 4.25 & 1.52 & 1.86 & 1.87 & 0.89\\
F & 6.85   & 175  & 1.31 & 1.49 & 2.30 & 0.93\\
G & 18.4   & 130  & 1.17 & 1.24 & 2.94 & 0.96\\
H & 61.4   & 155  & 1.08 & 1.10 & 4.00 & 0.98\\
I & 127    & 188  & 1.06 & 1.06 & 4.84 & 0.99\\
\hline
\end{tabular}
\end{center}
\medskip
$\dot{J}=\LL\FF$ and $\dot{H}=\HH\FF$. Values of $\FF$ are given in code units, 
see Eqs.~(\ref{eq:c1})-(\ref{eq:cf}) for conversion to physical units. In all
 cases $\Omega=0.143$.

\end{table}

\begin{figure}
\resizebox{\hsize}{!}{\includegraphics[clip=true]{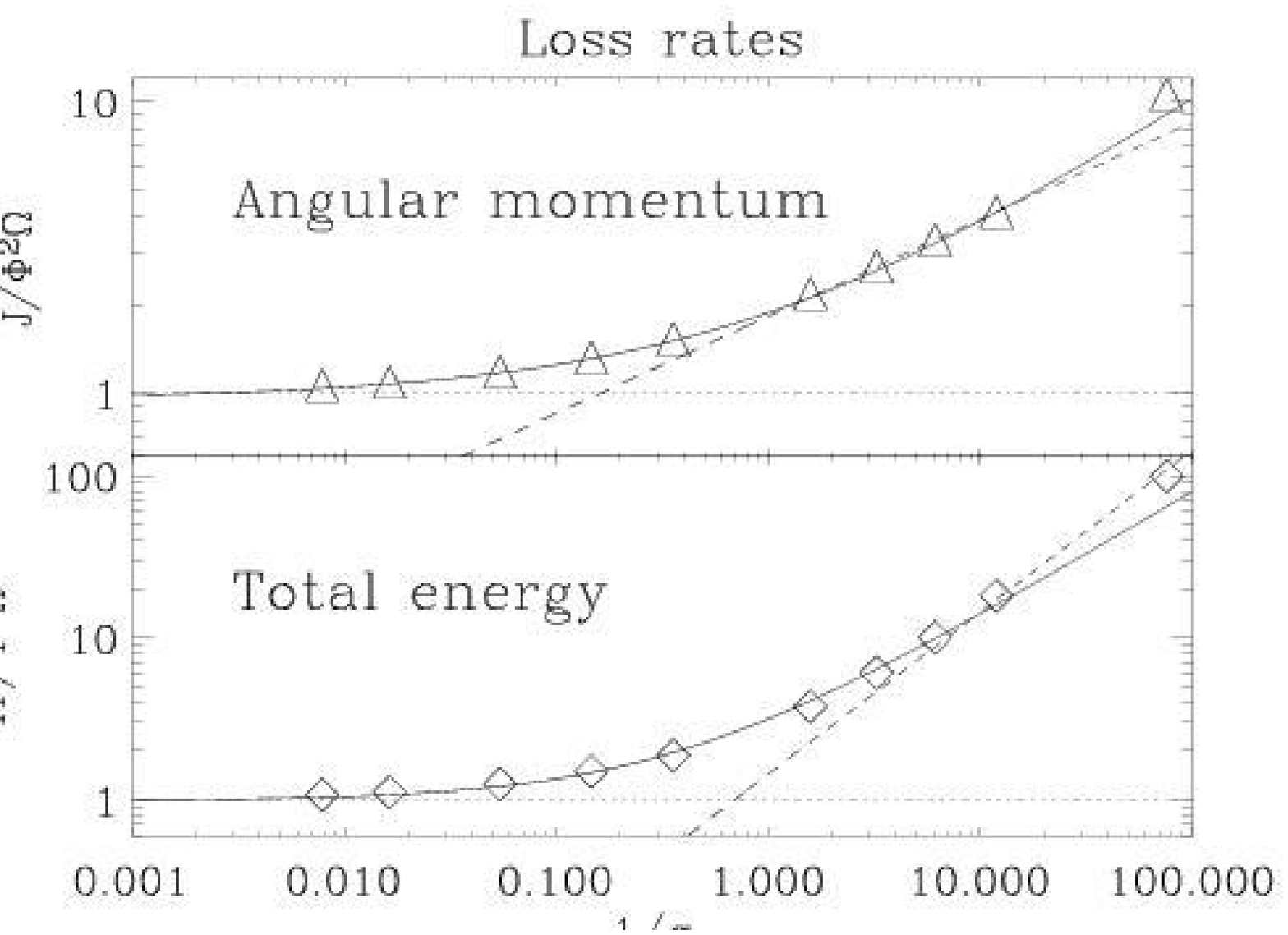}}
\caption{Loss rates for the 1D monopole calculations in
non-dimensional units (Table \ref{table:2}). Upper panel: angular
momentum loss rate. Lower panel: total energy loss rate. Dashed curves
represent the theoretical expectation for the losses in the mass
loaded cases $\dot{J}\propto \dot{M}^{1/3}$ and $\dot{H}\propto
\dot{M}$. Continuous curves represent the best power-low fit given in
the text. Dotted lines are the force-free solution.}
\label{fig:2}
\end{figure}

In Fig.~\ref{fig:1} we plot the velocity profiles for cases F, G, H,
and I, together with the location of the SM, AL, and FM points. Note
that the position of the SM point does not change significantly and is
given roughly by Eq.~(\ref{rsonic}). In Fig.~\ref{fig:2} the
angular momentum loss rate and energy loss rate are plotted as a
function of $\sigma_0$. The convergence to the force-free solution is
evident (see also Table~\ref{table:2}). An alternative way to parametrize
how close the soution is to the force-free limit is by considering
$r_A/R_L$ (\citealt{daigne02}), as shown in Tab.\ref{table:2}. In the mass-loaded cases
($\sigma_0<1$; models A$-$D) we find that $\dot{J}/\Phi^2\Omega\propto
\dot{M}^{1/3}$ in accord with the expectations from
\S\ref{sec:massloaded}.  We find that all the points (we excluded
case A0 because we are interested in case where $r_A>r_{SM}$,
typical of the transition from mass loaded to force-free) can be
approximated by a relation of the form
$\dot{J}/\Phi^2\Omega=C_o+C_1(1/\sigma_0)^{1/2}$, where $C_o$ should
correspond to the force-free limit $\dot{J}/\Phi^2\Omega=1$. A fit to
our results gives $C_o=0.98$ and $C_1=0.93$. A similar expression can
be written for the energy losses. At low $\sigma_0$ the total energy
loss rate scales as the mass loss rate, as expected,
$\dot{H}/\Phi^2\Omega^2\propto\dot{M}$. For large $\sigma_0$, 
the solution converges to the
force-free limit, $\dot{H}/\Phi^2\Omega^2=1$ (see Fig.~\ref{fig:2}). 
We find that the transition between the two limits can be fit with
$\dot{H}/\Phi^2\Omega=C_o+C_1(1/\sigma_0)^{0.77}$, where $C_o=0.98$
and $C_1=2.16$. We also consider the reduced energy loss rate (the
difference between the total energy and the rest mass energy), and
find that this too can be approximated with a power law:
$(\dot{H}-\FF)/\Phi^2\Omega^2=C_o+C_1(1/\sigma_0)^{0.61}$,
with $C_o=0.98$ and $C_1=0.965$. We stress that these power law
relations have been determined by fitting the results of our
simulations. As such, there is no guarantee that these trends can be
extended outside the range we investigated, but they do correspond well to the estimates
given in \S\ref{sec:physmod}.

Our simulations focus on the region close to the neutron star and so
the problem of the acceleration of the outflow at large distances
cannot be properly addressed. However, as shown, for example, by
\citet{daigne02} the efficiency of conversion from
magnetic to kinetic energy in the strict monopole limit is very
low. Faster than radial divergence in the flux tubes is required after
the FM point to increase the acceleration significantly
(\S\ref{sec:poynting}). Within our computational domain in Case I we
find that the ratio of particle kinetic energy flux to electromagnetic
energy flux scales approximatively as $r^{1/3}$ and shows a tendency
towards saturation. At a radius of $300 r_{NS}$ in Case I, the ratio
is still less than 5\%.
 
\


\subsection{The 2D Monopole}   
\label{section:2dmonopole}

The 1D monopole discussed above does not take into account
deformations of the poloidal field lines by the moving plasma. As a
consequence, the conversion of magnetic energy into kinetic energy of
the accelerated wind is inefficient. To understand if and how
deviations from a strict monopole may affect the dynamics of the
outflow it is necessary to perform 2D simulations. We focus only on
the region close to the star, within $100-200\,r_{NS}$, and we
consider both mass-loaded ($\sigma_0<1$) and Poynting-flux dominated
($\sigma_0>1$) regimes (see Tables \ref{table:1} \& \ref{table:3}).
In contrast to the cases considered by \citet{bog01}, where
the mass flux was fixed at $r_{in}$ and a cold wind ($p=0$) was
assumed, here the mass flux is derived self-consistently, with pressure
at the base of the wind being the control parameter for the flow. 
Even if $c_T(r_{in}) \sim 0.1c$, the difference with the pressureless case is not 
trivial. For example the location of the FM point in the 1D monopole is at infinity if
$p=0$, so, in principle, one might expect a higher efficiency also in the 2D case.
More important, in our case, the velocity at the base is much smaller that $c$, 
so that collimation in the region close to the star could be more efficient.  

\begin{figure*}
\resizebox{\hsize}{!}{\includegraphics[clip=true]{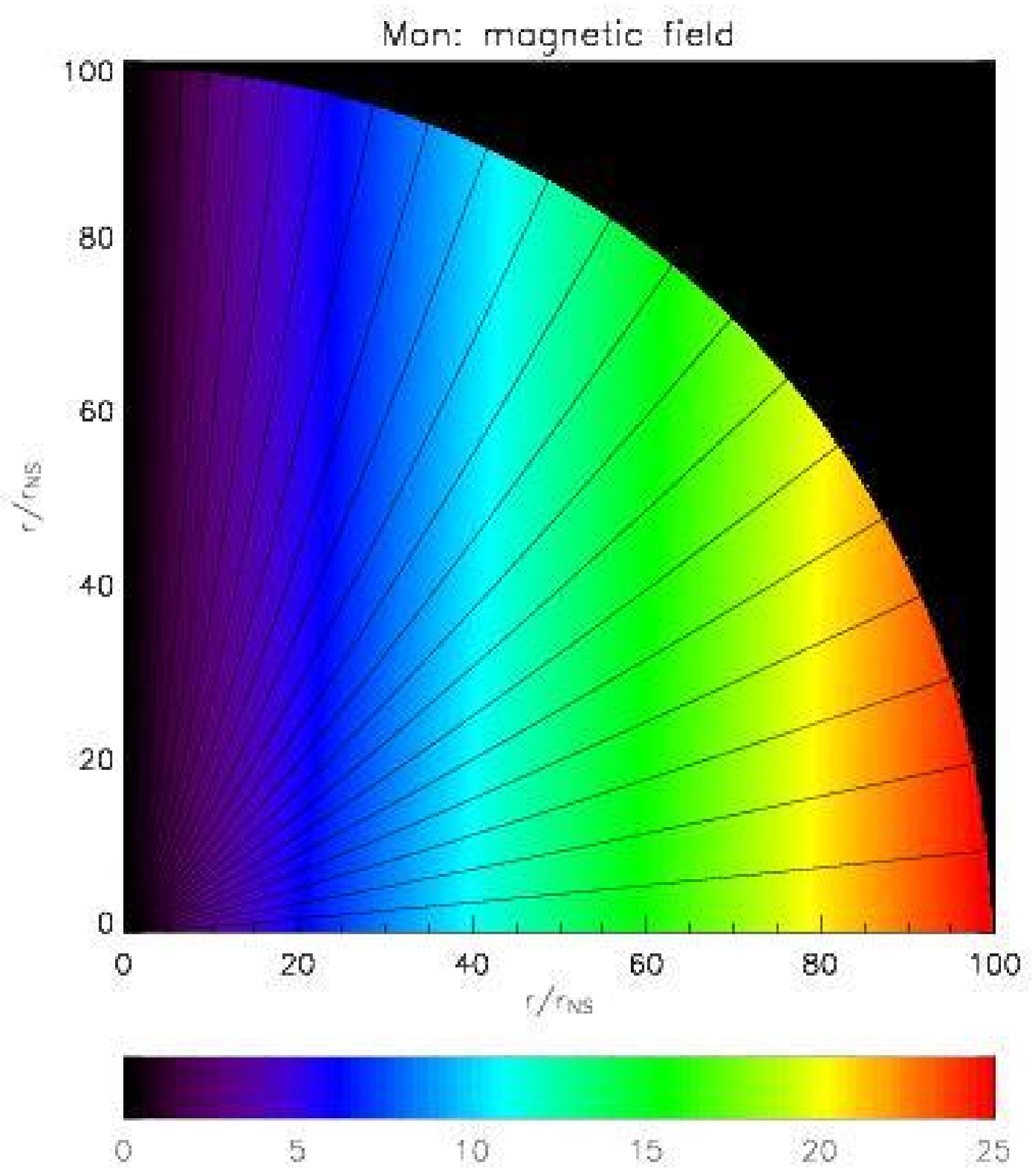}\includegraphics[clip=true]{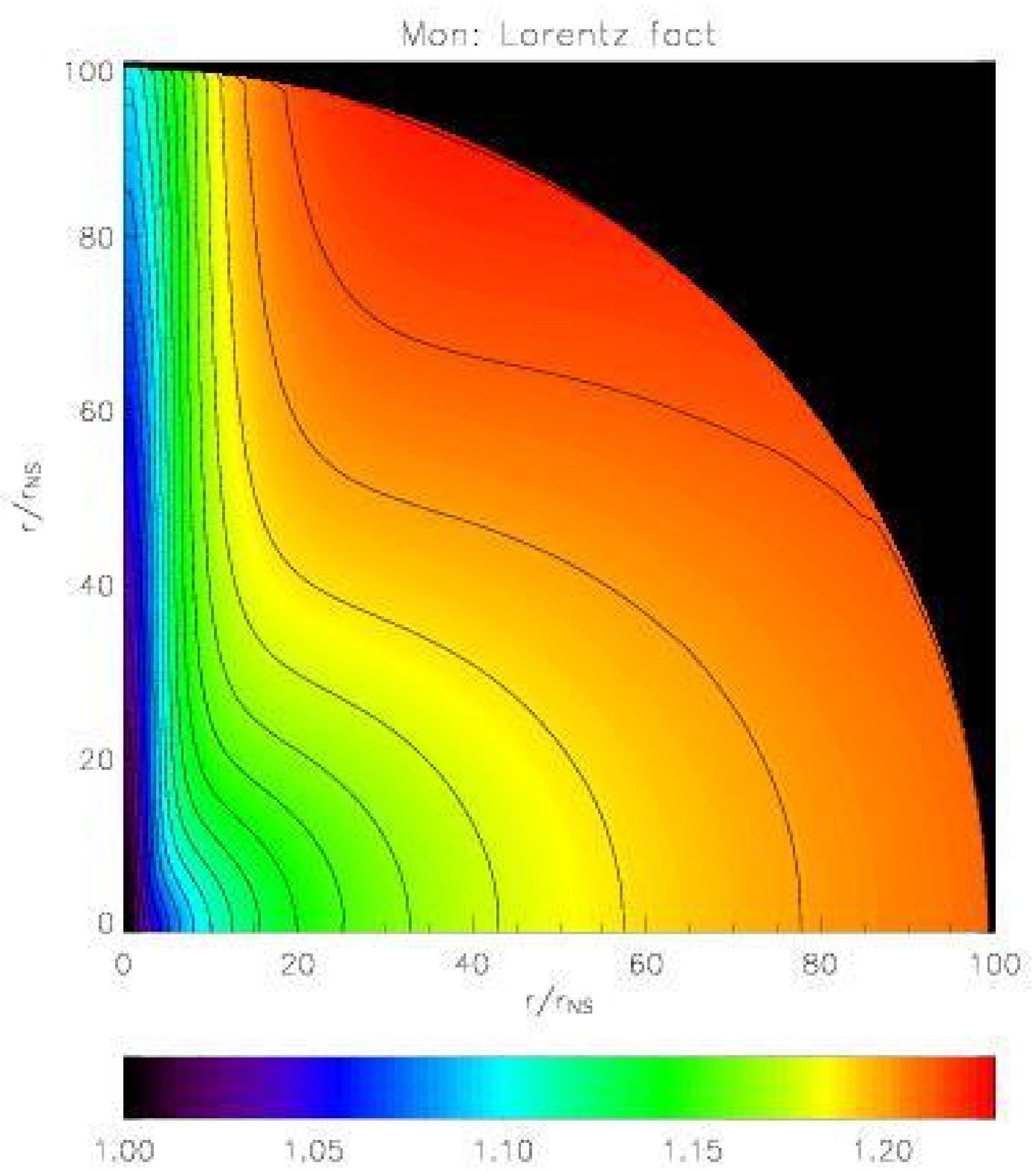}}
\resizebox{\hsize}{!}{\includegraphics[clip=true]{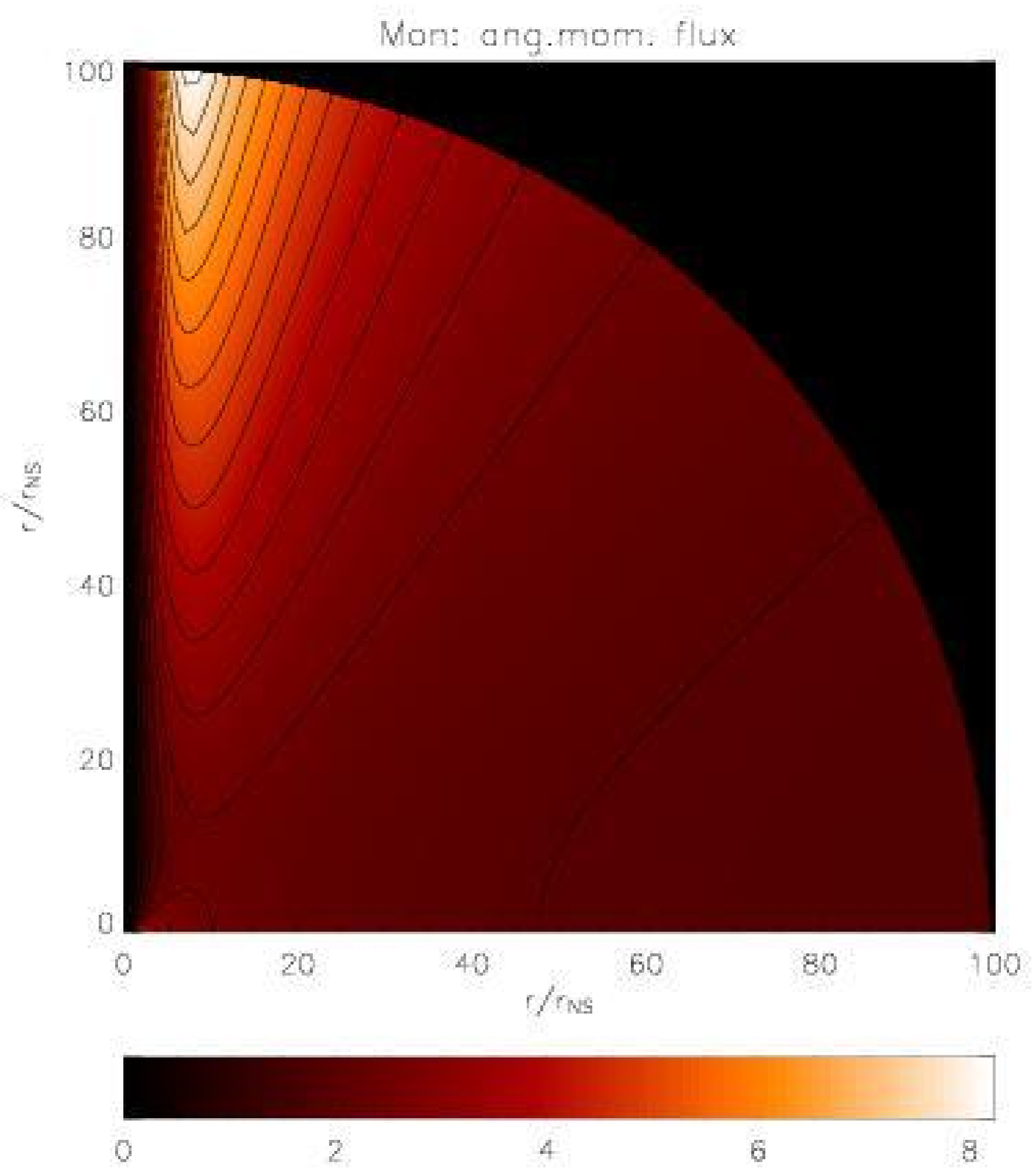}\includegraphics[clip=true]{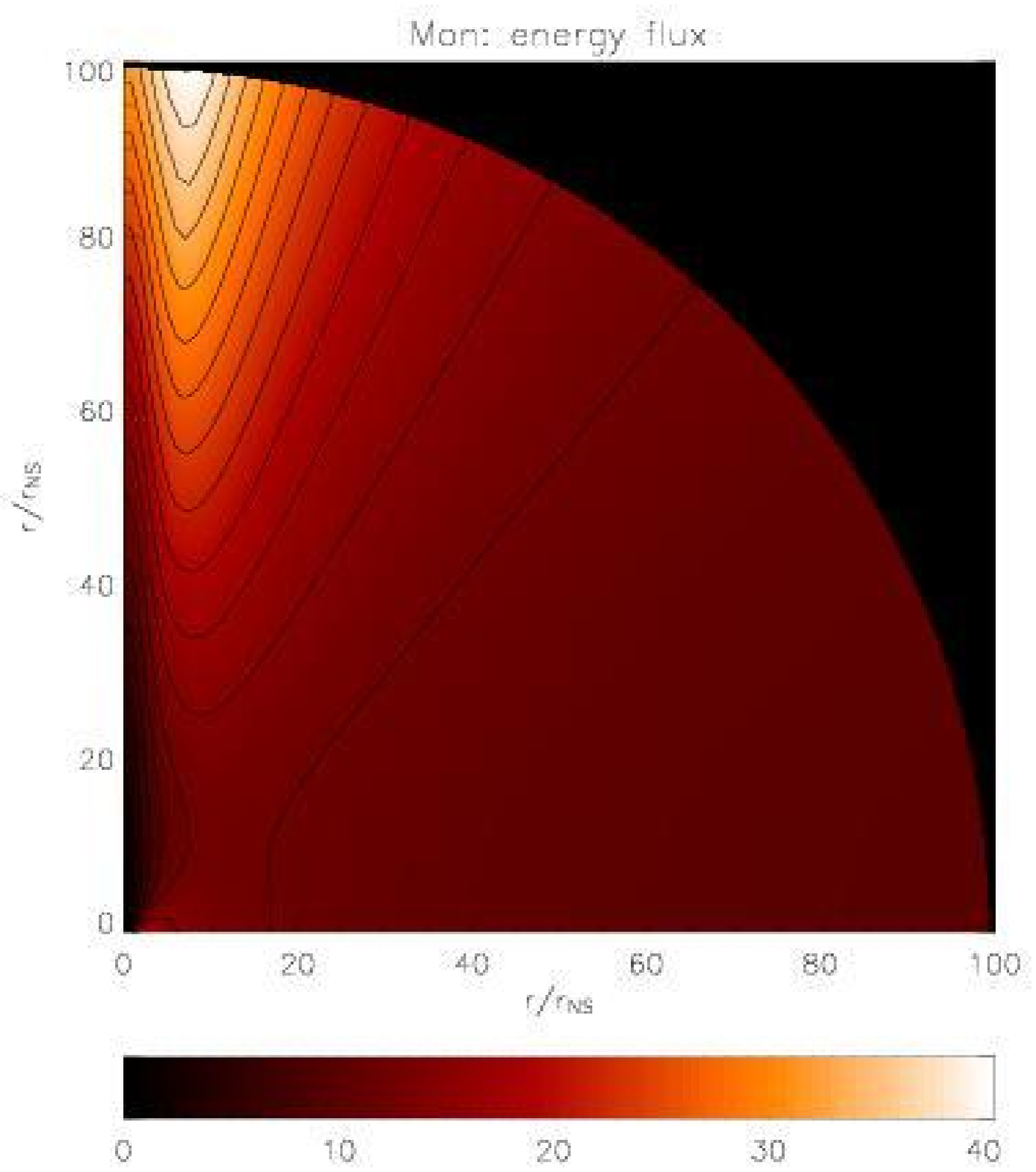}}
\caption{Results for the 2D monopole in the weakly magnetized Case A (Table \ref{table:3}).
 Upper left: contours represent poloidal magnetic field lines, while colors represent the 
ratio $|B_{\phi}/B_r|$. Upper right: colors and contour represent the Lorentz factor. Notice 
the presence of a slow channel on the axis and the peak in velocity at about $70^\circ$. 
 Lower left: angular momentum flux in adimensiona units $\LL\FF/(4\pi\Phi_t^2\Omega)$.
 Lower right: total energy flux $\HH\FF/(4\pi\Phi_t^2\Omega^2)$ in adimensional units
(see Eqs.~(\ref{eq:c1})-(\ref{eq:cf}) for conversion in physical units).
 Note that these fluxes peak at high latitudes (compare with 
Fig.~\ref{fig:4}).}
\label{fig:3}
\end{figure*}

\subsubsection{Magnetic, Mass-Loaded Winds}

In Fig.~\ref{fig:3} we show the results for a heavily mass-loaded case
(Case A) corresponding to a $\sigma_0 =
(4\pi)^{-1}\int_S\Phi^2\Omega^2/\FF\,ds=0.121$, where the integration
is performed over $4\pi$ solid angle.

\begin{table}
\caption{Results of the 2D monopole models.}
\label{table:3}
\begin{center}
\begin{tabular}{c c c c c c c}
\hline
Mod. & $\sigma_0$ & $\FF_{t}$ & $\dot{J}/\Phi_{t}^2\Omega$ & $\dot{H}/\Phi_{t}^2\Omega^2$ & $r_{A,equat}/R_L$\\
\hline
A & 0.121   & 98    & 1.23 & 11.9   & 0.40 \\
B & 1.00    & 119   & 1.11 & 2.28   & 0.68 \\
B1& 1.22    & 217   & 1.05 & 1.99   & 0.70 \\ 
C & 9.67    & 122   & 0.755 & 0.875 & 0.92 \\
D & 68.5    & 173   & 0.678 & 0.699 & 0.98 \\
E & 211.5   & 280   & 0.671 & 0.679 & 0.99 \\
\hline
\end{tabular}
\end{center}
\medskip

$\dot{J}=(4\pi)^{-1}\int_s\LL\FF ds$; 
$\dot{H}=(4\pi)^{-1}\int_s\HH\FF ds$; $\Phi_{t}=(4\pi)^{-1}\int_s\Phi ds$; 
$\FF_{t}=(4\pi)^{-1}\int_s\FF ds$ and $\sigma_0=\Phi_{t}^2\Omega^2/\FF_{t}$. Values 
are given in code units, see Eqs.~(\ref{eq:c1})-(\ref{eq:cf}) for conversion to physical 
units. The value of $\SH$ in all cases is 0.018. In all cases $\Omega=0.143$, except case B1 which has $\Omega=0.214$. $r_{A,equat}$ is the radial distance of the AL sulface on the equatorial plane.
\end{table}

 At $r_{in}$ the mass flux profile scales approximatively as $\sin^2{\theta}$, and
is minimal at the pole. As a result of magnetic acceleration and
centrifugal support at the equator, the mass flux is higher than in
the corresponding 1D hydrodynamical non-rotating case and it is about
the same as in the 1D monopole. However, at the pole the mass flux is
lower because of magnetic collimation on the axis (Kopp \& Holzer
1976).  The large difference in mass flux between pole and equator is
manifest in the elongated shape of the FM surface.  The upper left panel of
Fig.~\ref{fig:3} shows that the field lines are very close to radial
and that $B_\phi$ scales approximatively as $\sin{\theta}$. At the
pole, the FM surface falls outside of the computational domain,
whereas the FM surface intersects the equatorial plane at $5.2 r_{NS}$
($r_{in}=1.1 r_{NS}$).
 
The SM surface also has a large axis ratio:
it intersects the pole at a distance of $11 r_{NS}$ and it intersects the 
equator at $1.8 r_{NS}$ It is interesting to look also at the position of 
the AL surface. Its distance from the Light Cylinder $R_L$ is an indicator 
of how close the solution is to the force-free limit and it strongly 
reflects the degree of magnetization. While on the axis where $B_\phi=0$ 
the AL and FM surfaces are coincident, away from the pole the toroidal 
field component does not vanish and the two critical curves separate. 
The AL surface intersects the equator at a distance of about $2.7 r_{NS}$ 
(to be compared with $R_L=6.8r_{NS}$).  

One might expect the Lorentz factor to be largest in the equatorial
region, as a result of stronger magnetocentrifugal effects
there. However, contrary to this expectation, we observe that $\gamma$
peaks at about $70^{\circ}$. Such a result was also obtained by
\citet{bog01} for cold flows. This effect is stronger in our
calculations because of the lower overall Lorentz factor. We also
notice the existence of a very slow channel along the axis. We want to
stress that the FM surface is outside the computational domain within
3$^\circ$ of the axis, and so the solution has not converged fully in
this region. However, by increasing the radial computational domain,
we find that the main effect of failing to capture the FM surface at
the pole is that the wind in this region is {\it less} collimated and
somewhat {\it faster} than it should be. So we expect the wind to be
more collimated and slower, with yet larger computational
grids. Whether the FM surface is inside the computational domain at
the pole or not has relatively little effect on those streamlines at
lower latitudes that do pass through the FM point.  Typical deviations
are found to be less than 1\%.

As the lower panels of Fig.~\ref{fig:3} show, we find that both the
energy flux (which is mainly kinetic) and the angular momentum flux
peak at high latitudes. In addition, the mass flux at large distance
from the neutron star is higher close to the axis because of magnetic
collimation. In contrast, the conversion of electromagnetic energy to
kinetic energy is maximal along the equator, even though in this
mass-loaded low-$\sigma_0$ case the electromagnetic energy flux is lower
than kinetic energy flux, and the wind terminal Lorentz factor is
mainly given by the conversion of internal energy to kinetic energy.


\subsubsection{Poynting-Flux Dominated Winds}

\begin{figure*}
\resizebox{\hsize}{!}{\includegraphics[clip=true]{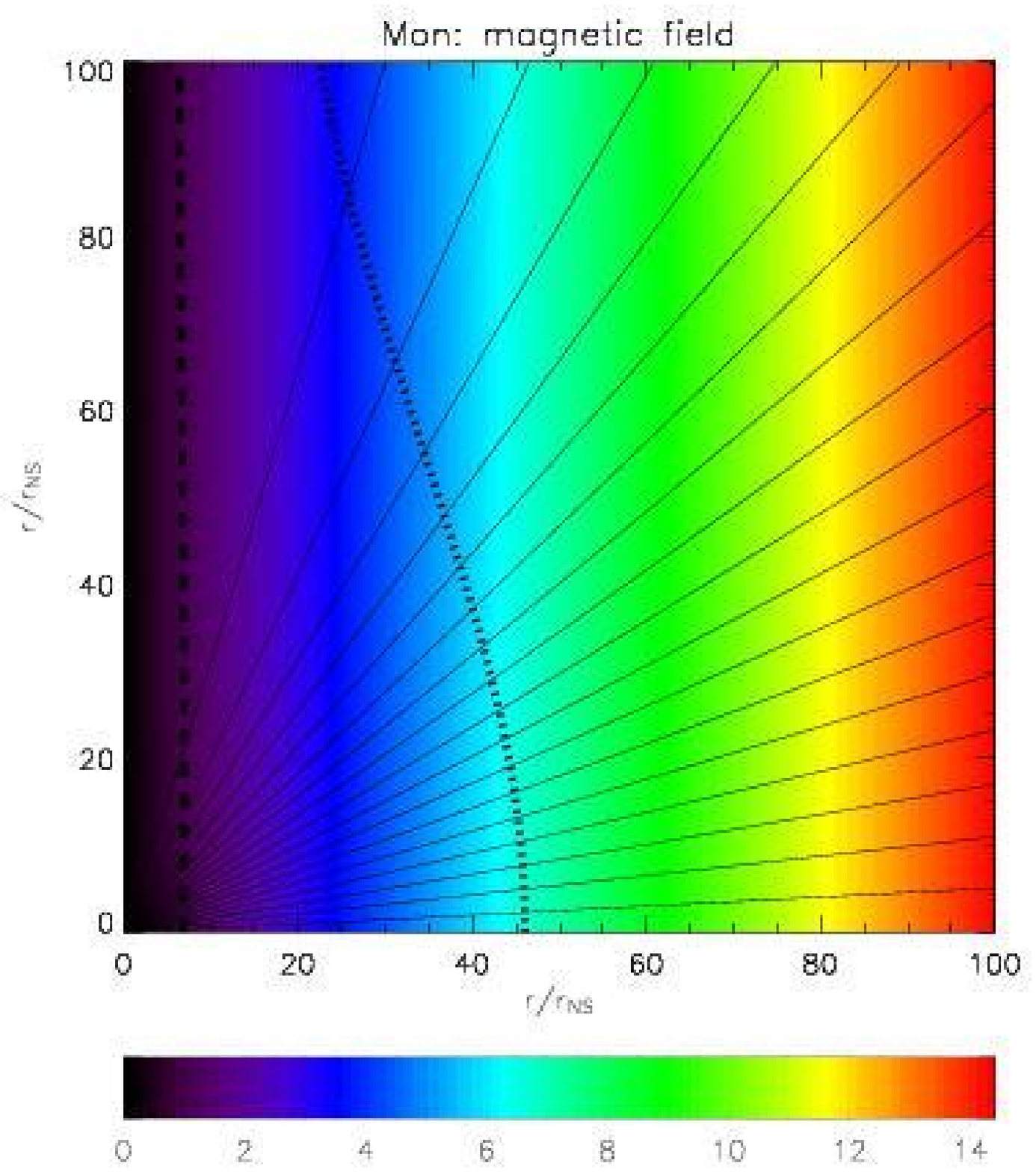}\includegraphics[clip=true]{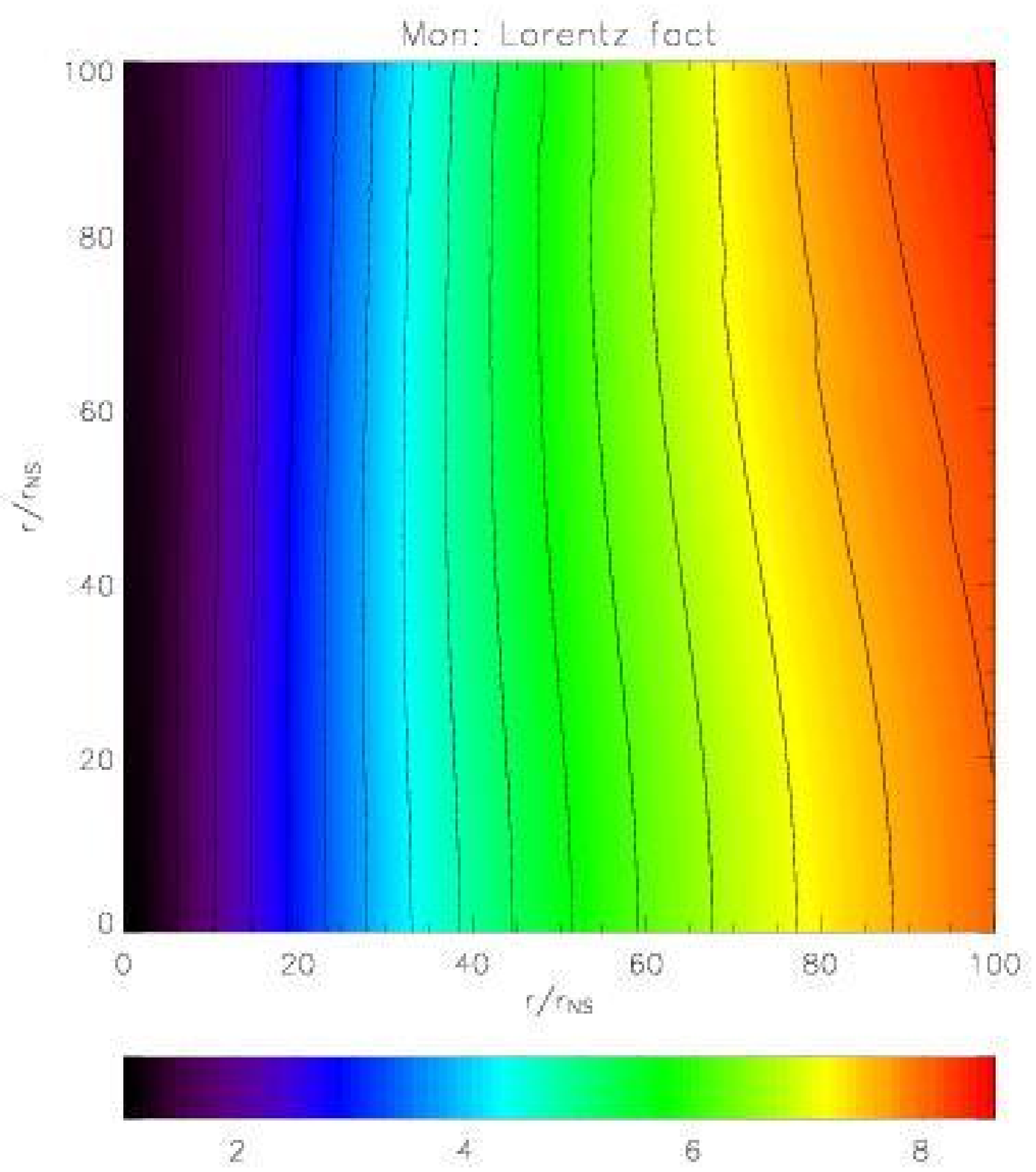}}
\resizebox{\hsize}{!}{\includegraphics[clip=true]{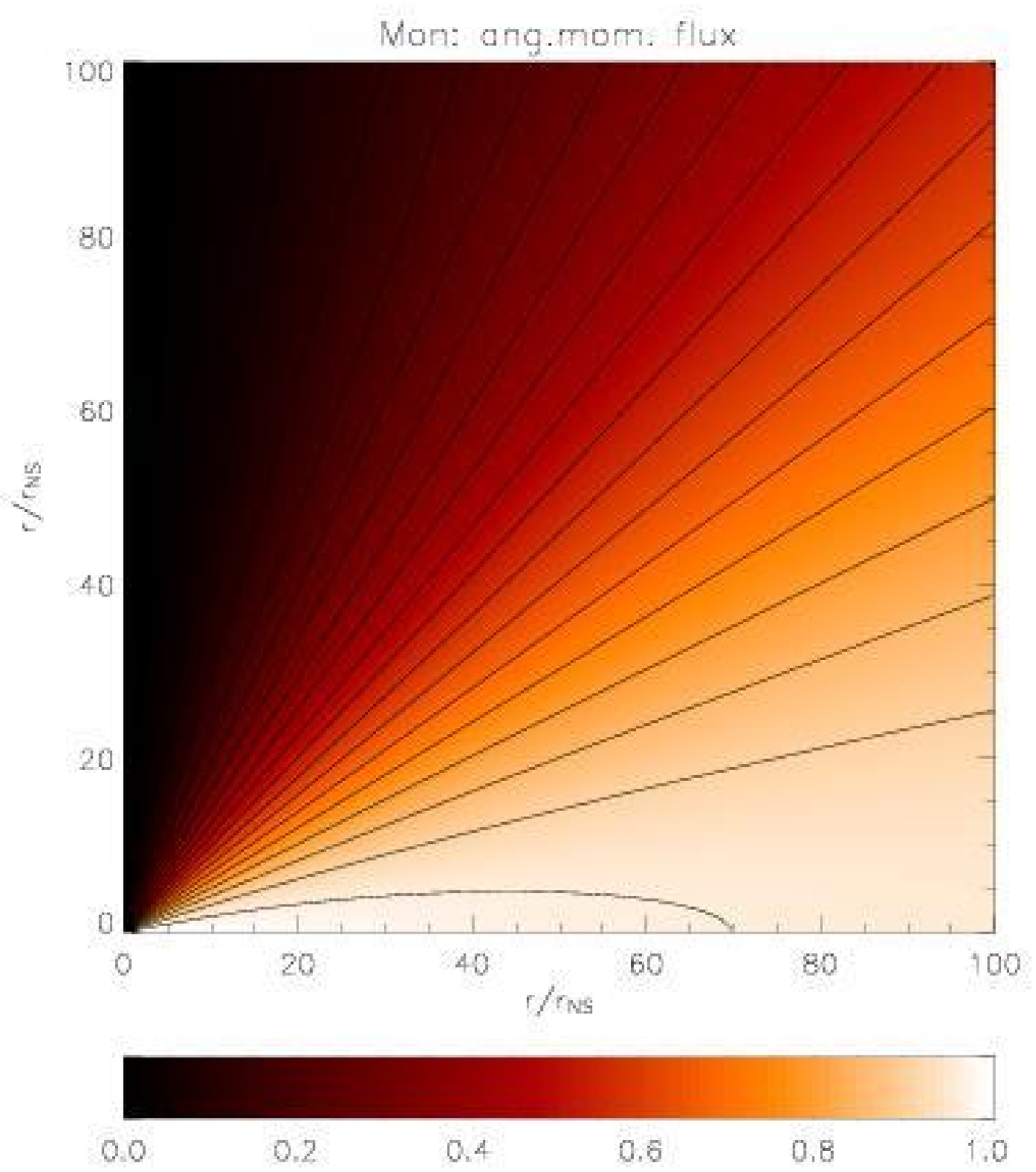}\includegraphics[clip=true]{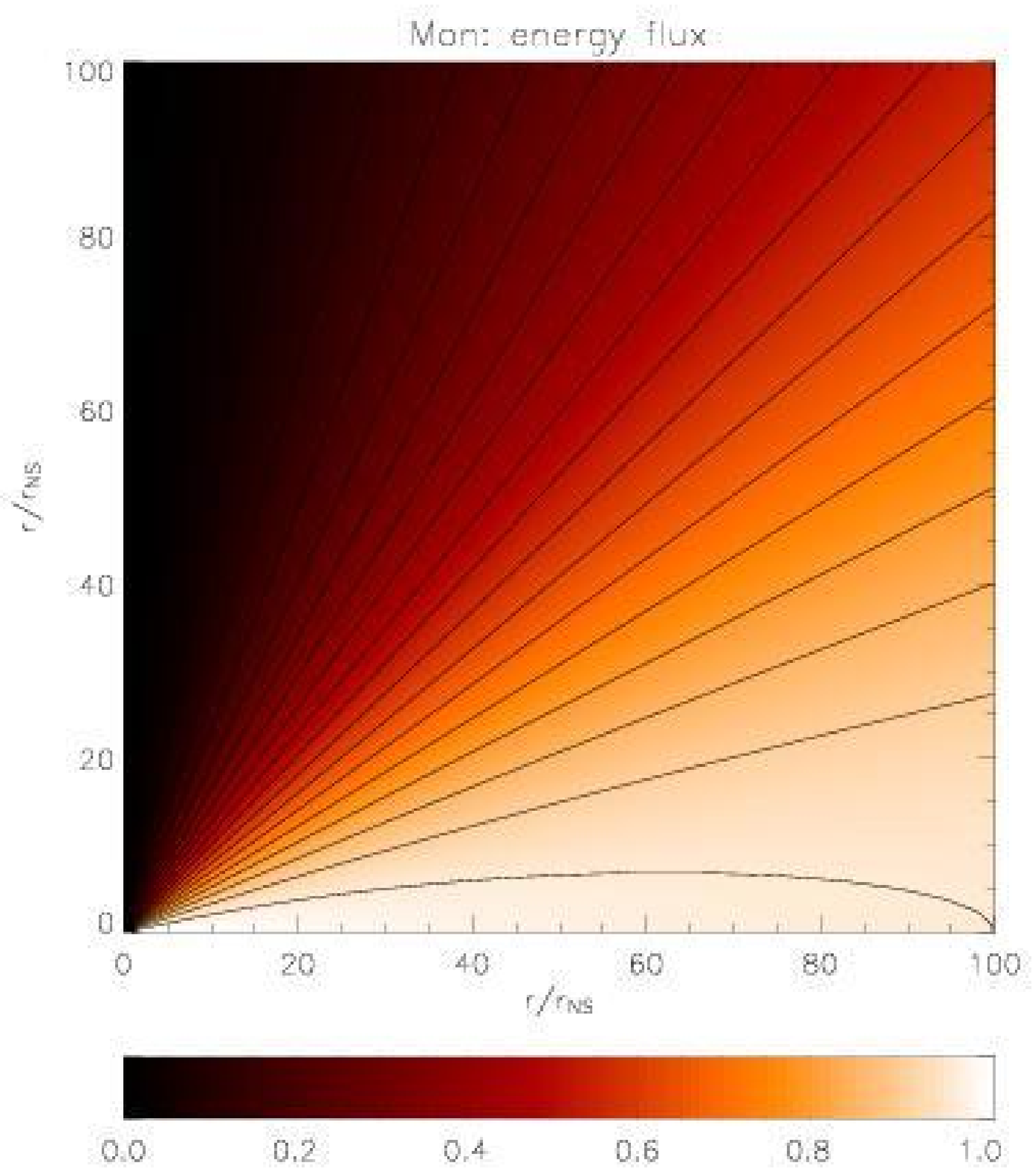}}
\caption{Results for the 2D monopole in the highly magnetized Case E (Table 
\ref{table:3}). Upper left: contours represent poloidal magnetic field lines,
 while colors represent the ratio $|B_{\phi}/B_r|$. The FM surface (dotted line) 
is  more distant from the axis while the AL surface (dashed line) is very close 
to the Light Cylinder. Upper right: colors and contours represent the Lorentz 
factor. There is no evidence here of the relatively slow channel on the axis as 
in Case A (see Fig.~\ref{fig:3}) and the Lorentz factor scales as $\sin{(\theta)}$.
 Lower left: angular momentum flux in adimensiona units $\LL\FF/(4\pi\Phi_t^2\Omega)$.
 Lower right: total energy flux $\HH\FF/(4\pi\Phi_t^2\Omega^2)$ in adimensional units
(see Eqs.~(\ref{eq:c1})-(\ref{eq:cf}) for conversion in physical units).
 Here, these fluxes are higher at the equator
 than at the pole (compare with Fig.~\ref{fig:3}), as expected when the flow 
is relativistic and Poynting-flux dominated.}
\label{fig:4}
\end{figure*}

In Fig.~\ref{fig:4} we show the results for a Poynting-flux dominated
flow (Case E), corresponding to a ratio $\sigma_0 =
(4\pi)^{-1}\int_S\Phi^2\Omega^2/\FF\,ds=211.5$, where the integration
is again performed over $4\pi$ solid angle. In this case the solution
is mostly magnetically driven, and plasma effects lead to small
deviations with respect to the force-free limit.

Similar to the mass-loaded Case A, at the neutron star surface we find 
that the mass flux is higher at the equator than at the pole, and that 
the AL and FM surfaces are extended in the direction of the pole. 
Here, the FM and AL surfaces intersect the equatorial plane at a
distance of $46 r_{NS}$ (to be compared with $40.5=\sigma_0^{1/3}R_L$)
 and $6.7 r_{NS}$, respectively, while the SM
surface is more spherical than Case A and intersects the pole and the
equator at a distance of $2.5 r_{NS}$ and $1.7 r_{NS}$,
respectively. Magnetic field lines are again monopolar, but while in
the mass-loaded Case A this was a consequence of a amrginally centrifugally driven
wind, now this is due to electromagnetic force balance. The dynamics
of a magnetized outflow are governed by the combination of Lorentz and
Coulomb forces. In the relativistic regime the Coulomb force cannot be
neglected and, as the flow speed approaches $c$, the Coulomb force
balances the Lorentz force, suppressing collimation. As a consequence,
the flux tubes in a Poynting-flux dominated wind have an areal
cross-section that scales as $r^2$. It is known that the efficiency of
conversion of magnetic energy to kinetic energy increases when the
flow divergence becomes more than radial \citep{daigne02}. In the
high-$\sigma_0$ simulation presented here, at least within the limited
computational domain employed, we do not see evidence for efficient
conversion and acceleration. As found by \citep{bog01} there is evidence
for a narrow collimated channel very close to the axis, but in our case, where the
mass flux at the injection is not imposed, the mass flux in the channel is 
not strongly enhanced. However the FM surface on the axis does not close 
in our computational box so no strong conclusion can be drawn.

The upper right panel of Fig.~\ref{fig:4} shows that the Lorentz
factor in the wind scales approximatively as $\sin{\theta}$.  The
maximum value achieved in the computational domain is 8. The latitude
dependence is not appreciable, probably because our solution does not
extend far enough away from the star. As in the 1D models, for the
range of parameters investigated, the total mass flux from the star is
not much affected by the value of the surface magnetic field.  For the
Poynting-flux dominated Case E it is about 3 times higher than in the
mass-loaded Case A, despite the fact that the magnetic energy density
is 5000 times higher. This again follows from the fact that the AL
surface is larger than the SM surface.  However, there are important
qualitative differences between Case A and Case E.  In case E we find
that on large scales magnetic acceleration is dominant and the mass
flux is maximal on the equator, instead of on the axis. In addition,
the lower panels of Fig.~\ref{fig:4} show that the energy and angular
momentum fluxes increase toward the equatorial plane and are almost
completely magnetically-dominated. In fact, the energy and angular
momentum loss rates scale as $\sin^2{\theta}$, as expected in the
force-free limit.  This strong transition in the angular dependence of
the energy and momentum fluxes, from strongly collimated to
equatorial, occurs at $\sigma_0\sim1$.

\begin{figure}
\resizebox{\hsize}{!}{\includegraphics[clip=true]{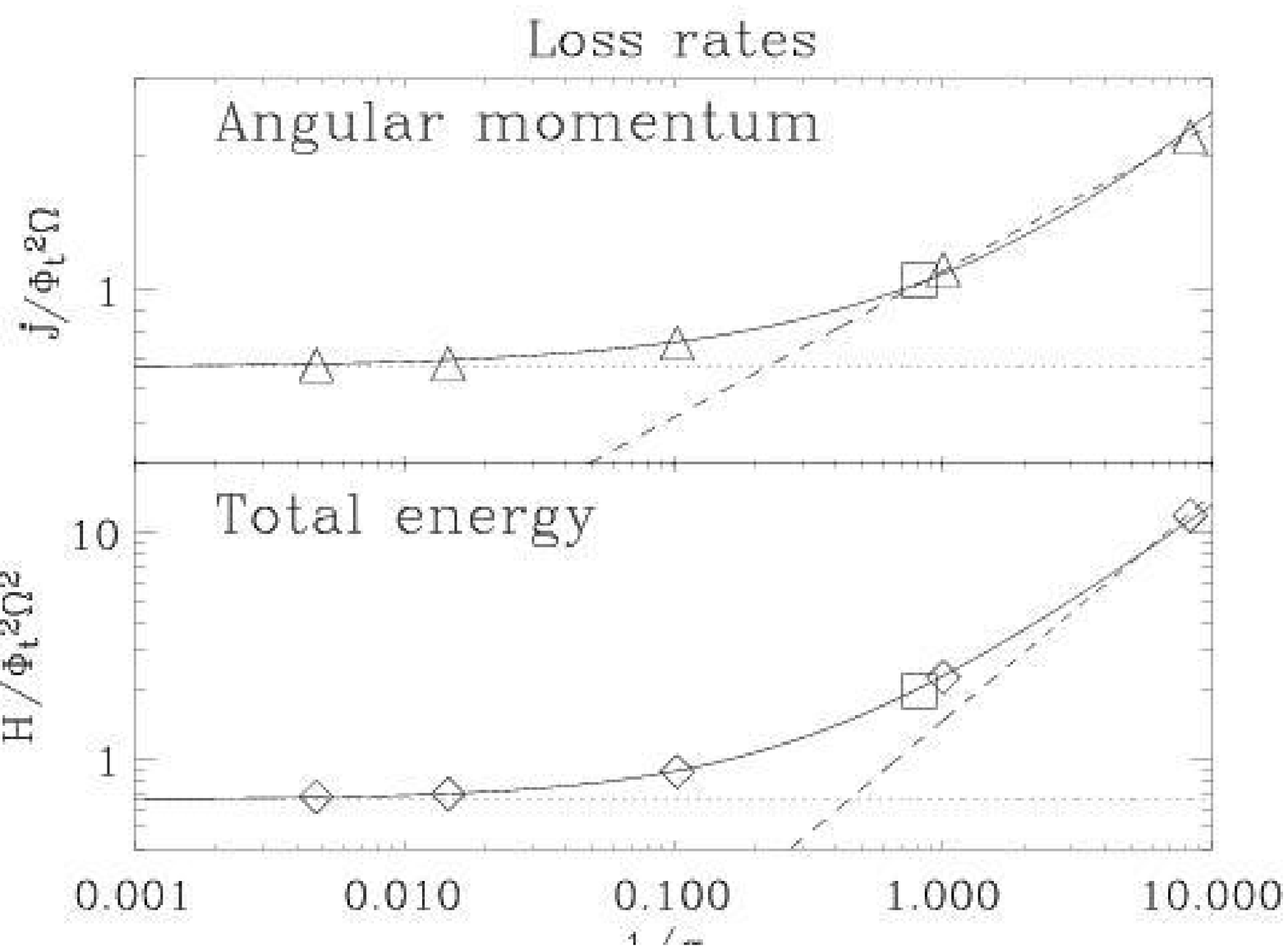}}
\caption{Loss rates for a 2D monopole in non-dimensional units. Upper
panel: angular momentum loss rate. Lower panel: total energy loss
rate. Dashed curves represent the theoretical expectation for the
losses in the mass loaded cases $\dot{J}\propto \dot{M}^{1/3}$
(\S\ref{sec:massloaded}) and $\dot{H}\propto \dot{M}$. Continuous
curves represent the best power-low fit given in the text. The dotted
lines are the force-free limits. The square mark indicates Case B1,
which has a different rotation rate (Tab.~\ref{table:1}).}
\label{fig:5}
\end{figure}

Figure \ref{fig:5} shows the behavior of the angular momentum and
energy losses as $\sigma_0$ increases. We again observe the convergence
to the force-free limit (see also Tab.~\ref{table:3}) and recover the
expected behavior $\dot{J}/\Phi^2\Omega\propto \dot{M}^{1/3}$ in the
mass-loaded cases. The convergence to the force-free solution can be
approximated, as in the 1D case, with a power law of the form
$C_o+C_1(1/\sigma_0)^{\beta}$. The fits are different than the 1D
models because the flux tubes deviate from being strictly
monopolar. We find for the angular momentum loss rate
$\dot{J}/\Phi^2\Omega=0.664+0.42(1/\sigma_0)^{0.67}$, for the
total energy losses
$\dot{H}/\Phi^2\Omega^2=0.660+1.66(1/\sigma_0)^{0.87}$, and
for the reduced energy
$(\dot{H}-\FF)/\Phi^2\Omega^2=0.66+0.60(1/\sigma_0)^{0.74}$. As
mentioned before, we have derived these fits within the parameter
range of our simulations, and there is no guarantee that they can be
extrapolated for much higher or lower magnetizations. We stress that
in the highly-magnetized Case E our results are very close to the
force-free solution: $\dot{J}/\Phi^2\Omega=\dot{H}/\Phi^2\Omega^2=2/3$
\citep{michel91}. Note also that such a value is lower than what would be
expected from a trivial 2D extension of the 1D monopole
(0.78=$\int_o^{\pi/2}\sin{(\theta)}^2\,d\theta$).

The efficiency of conversion of the magnetic energy to kinetic energy
is maximal on the equator, but it does not exceed 10\% at the outer
boundary. We note that conversion is faster than logarithmic in radius
and at the edge of the computational box it scales as $R^{1/3}$,
similar to our 1D results. However, we cannot draw strong conclusions
on the terminal efficiency far outside of the FM surface because of
the limited size of our computational domain. As shown by 
\cite{bog01}, $\gamma$ seems to increase after the FM point
and then saturates at larger distances.

\subsection{The Aligned Dipole}
\label{sec:dipole}

To lowest order, currents in the neutron star should generate a dipole
magnetic field.  This field configuration is much more realistic than
the monopolar models considered in \S\ref{section:1dmonopole} and
\S\ref{section:2dmonopole}.  A dipole field may also have interesting
consequences for the asymptotic character of the outflow.  For
example, it is possible that the presence of a closed zone, outside of
which the open field lines at first expand much more rapidly than radially, might
provide for a more efficient conversion of magnetic energy into
kinetic energy, leading to a higher terminal Lorentz factor (\S\ref{section:introduction}).  
Below, we review several numerical issues associated with our dipole wind
solutions and then we present our results, which again bridge the
transition from low- to high-$\sigma_0$ outflows.


\subsubsection{Numerical Challenges}

The modeling of a magnetic wind with a closed zone and an equatorial current 
sheet presents a number of numerical difficulties. We have encountered two
 problems in particular that bear mention.  The first is that the outer 
edge of the closed zone rotates {\it faster} than what is required by 
Eq.~(\ref{eq:rot}). We believe this ``super-corotation'' is connected with 
numerical dissipation in the equation for the evolution of $B_\phi$ and 
that it stems from the fact that the boundary between the closed and open
field regions is not grid-aligned.\footnote{Increasing the resolution of
the simulation or using a characteristics-based solver (\citealt{komissarov99},
private communication) does not improve the accuracy of the solution.}
By suppressing the upwinding term in the HLL flux we were able to reduce 
the deviation from the corotation condition from $10-20$\% to $\sim$5\%,
but at the price of making the code less stable when the flow is highly
magnetized. Unfortunately, of the previous papers dealing with winds in
the presence of a closed zone in the MHD regime, only
\citet{kep00} discuss deviations from corotation in the closed zone.
In their paper these amount to $\sim$10-20\%, and they consider only 
parameters appropriate to the Sun, a slow rotator.

\begin{figure}
\hspace*{-0.7cm}\includegraphics[width=11cm,height=9cm,clip=true]{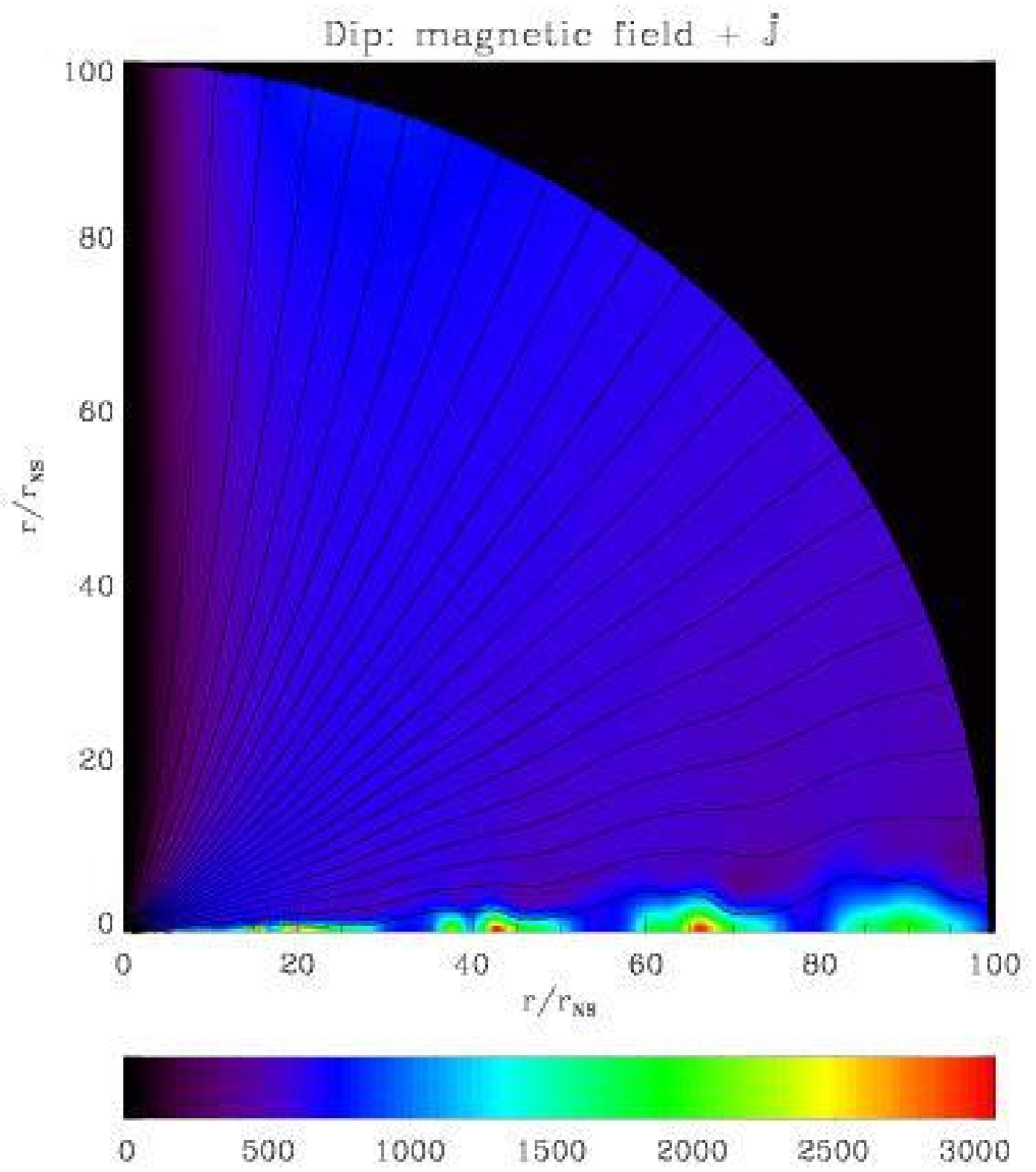}
\hspace*{-.7cm}\includegraphics[width=11cm,height=9cm,clip=true]{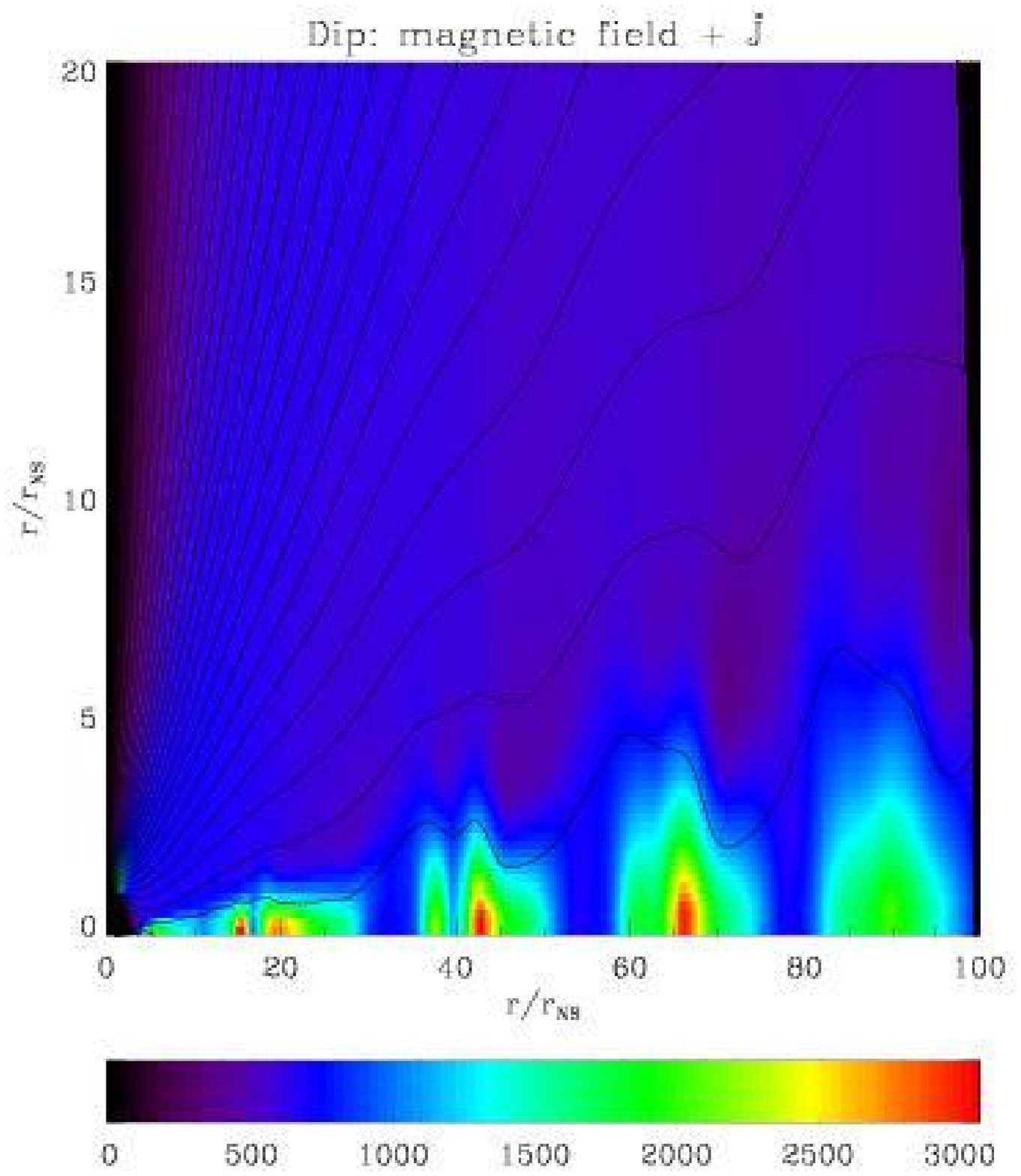}
\caption{Magnetic structure of a magnetically dominated flow (Case C). 
The upper panel shows the poloidal magnetic field structure, with a snapshot of 
outflowing plasmoids forming along the equatorial current sheet outside 
the closed zone. The lower panel shows a blow up of these plasmoids, 
which travel out at close to the speed of light.  The color indicates 
the angular momentum density, with roughly 20\% of the angular momentum 
loss being carried by these intermittent structures. Values are in code units
(Eqs.~(\ref{eq:c1})-(\ref{eq:cf})). As stated in the text, the numerical values
in the plasmoids depend on numerical resistivity.}
\label{fig:plasmoid}
\end{figure}

The second problem is that the magnetic field undergoes reconnection
at the neutral current sheet on the equator close to the position of
the $Y$ point, where the last closed field line intersects the
equatorial plane. As a consequence, plasmoids are formed and advected
away, thus preventing the system from reaching a steady-state
configuration.\footnote{Wind calculations performed over a full 180
degrees show that the formation of plasmoids is not an artifact of our
90$^\circ$ computational domain.}  At the current sheet $B_r$ and
$B_\phi$ change sign, the MHD approximation fails, and a sharp
discontinuity develops that cannot be well-resolved. Figure
\ref{fig:plasmoid} shows an example of plasmoid formation.
Although it is well known that current sheets in the presence of a $Y$-type point 
are subject to reconnection and the continuous formation of plasmoids 
(\citealt{endeve03}, \citealt{tanuma05}, \citealt{yin00}), in our 
case the high value of $\sigma_0$ does not allow us to properly resolve 
the current sheet. For this reason, the reconnection processes are
dominated by the intrinsic numerical resistivity of our numerical scheme.

\begin{table}
\caption{Results of the dipole models.}
\label{table:4}
\begin{center}
\begin{tabular}{c c c c c c c}
\hline
Mod. & $\sigma_0$ & $\FF_{t}$ & $\dot{J}/\Phi_{t}^2\Omega$ & $\dot{H}/\Phi_{t}^2\Omega^2$ & $r_{Y}/R_{L}$ \\
\hline
A & 0.298    & 66   & 1.53  & 5.28  & 0.26 \\
B2& 1.34     & 7.8  & 0.965 & 1.81  & 0.32 \\
B & 2.77     & 43   & 0.853 & 1.27  & 0.37 \\
B1& 3.91     & 114  & 0.816 & 1.10  & 0.48 \\
C & 17.5     & 42   & 0.701 & 0.73  & 0.47 \\
\hline
\end{tabular}
\end{center}
\medskip

 $\dot{J}=(4\pi)^{-1}\int_s\LL\FF ds$; $\dot{H}=(4\pi)^{-1}\int_s\HH\FF ds$; 
$\Phi_{t}=(4\pi)^{-1}\int_s\Phi_o ds$; $\FF_{t}=(4\pi)^{-1}\int_s\FF ds$ and 
$\sigma_0=\Phi_{t}^2\Omega^2/\FF_{t}$. Values are given in code units, see 
Eqs.~(\ref{eq:c1})-(\ref{eq:cf}) for conversion to physical units. Cases B1 
and B2 have different rotation rates (Tab.~\ref{table:1}). All cases have 
$\SH=0.018$.
\end{table}

We have found that the formation and growth of plasmoids depends on
two terms in the definition of the HLL electric field (see Eq.~44 of
\citealt{ldz03}): $\partial B_{r}/\partial \theta$ and
$B_{r}v_{\theta}$. The first behaves like an explicit resistivity and
seems responsible for the evolution of the plasmoids as they are
advected off the grid along the equator.  The second term behaves like
forced reconnection and controls the initiation of plasmoid
formation (given that the current sheet is not resolved, $v_{\theta}$ is not 
reconstructed to zero on the equator). Setting both terms to zero causes the closed zone to
disappear entirely and the system evolves toward a modified split
monopole.  To deal with this issue and explicitly enforce a steady
state, we opt for the following procedure: in a calculation with
plasmoids we note the position of the $Y$ point and we then impose the
conditions $\partial B_{r}/\partial \theta=0$ and $B_rv_\theta=0$ on
the equator outside the position of the $Y$ point as inferred from the calculation
without these boundary conditions. The condition $\partial B_{r}/\partial \theta=0$ 
can be justified because of the structure of the characteristic waves and because 
the solution should be symmetric about the equatorial plane. The equator is a contact 
discontinuity, so only the total pressure is important as a boundary 
conditions and not the sign of the parallel magnetic field, in fact for infinite 
conductivity a pure monopole and a split monopole have the same solution. The condition 
$B_rv_\theta=0$ corresponds to enforcing $v_{\theta}=0$.

We note that the steady state solution need not be the physically
correct solution found in nature.  The equatorial current sheet in the
vicinity of a neutron star is undoubtedly dissipative, and is thus subject
to reconnection and plasmoid formation.  However, a full understanding of this 
behavior requires --- at the very least --- the use of resistive RMHD,
so that the dissipation can be controlled, rather than being fully numerical, 
as in our current calculations.  Such a treatment is beyond the scope of this paper.
Thus, we focus our attention here on the forced steady state solutions
described above.  As a test, we have compared the global energy and
angular momentum loss rates between time-dependent calculations with
plasmoids and those with our forced steady state boundary conditions
outside the $Y$ point. In general, losses are higher in the latter
calculations because the open magnetic flux is larger.  In Case A the
difference is less than 5\%, while in Case C it is about 15-20\% (see
Tab.~\ref{table:4} and Fig.~\ref{fig:6}).  In the time-dependent
calculations the individual plasmoids represent fractional deviations
in $\dot{J}$ and $\dot{H}$ from the average of up to 15\% in Case C,
and less for lower $\sigma_0$ flows.  Thus, the plasmoids do not appear
to be that dynamically significant for the overall energy and angular
momentum losses from the neutron star.


\subsubsection{Results}

In Table \ref{table:4} we show the results of our simulations using
the prescription described above. In the monopole models the magnetic
field lines are open so that normalizing to the surface magnetic field
or to the total magnetic flux $\Phi$ is equivalent. In the dipole
case, because of the closed zone, the two quantities are not
proportional. We find it useful to normalize in terms of the magnetic
flux evaluated on open field lines, $\Phi_o$, where equations
(\ref{eq:integi})-(\ref{eq:integf}) hold. As a consequence, we can
define an equivalent surface magnetic field $B_{r-equiv}(r_{in})$ as
the surface magnetic field of a monopole that has the same amount of
open magnetic flux. Contrary to the monopole case, increasing the
magnetic field strength at the stellar surface {\it reduces} the mass
loss rate because the size of the closed zone increases. Similarly,
for the parameters explored here, an increase in the magnetic field
strength by a factor of, say, $\sim3$ at the stellar surface leads to
an increase in the open magnetic flux $\Phi_o$ by a factor of just
$\sim2.5$, rather than the one-to-one behavior for the monopole.
We can derive an approximate relation between the magnetic field at the pole 
on the surface of the star ($r_{NS}$), and the equivalent surface magnetic field,
 that can easily be computed if $\dot{H}$ and $\Omega$ are known. We find that 
in all our cases the relation is
\be
B_r(r_{NS},\theta=0)\simeq 1.6\, B_{r-equiv}(r_{NS})\times(r_Y/r_{NS}).
\label{equiv}
\ee 
Thus, the problem of relating the surface magnetic field to the loss rates is 
reduced to the problem of determining the size of the closed zone.

\begin{figure}
\resizebox{\hsize}{!}{\includegraphics[clip=true]{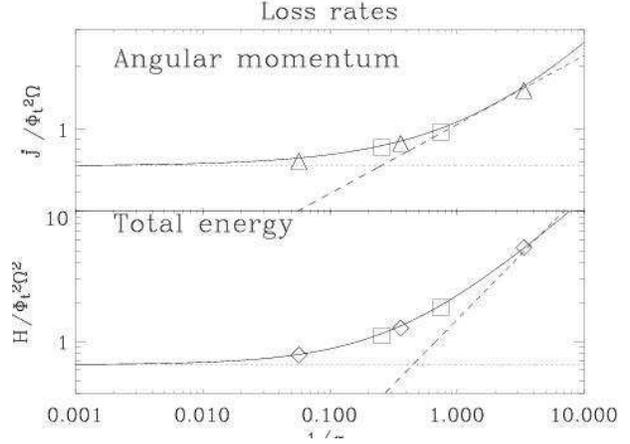}}
\caption{Loss rates for a 2D dipole in non-dimensional
units. Upper panel: angular momentum loss rate. Lower panel: total
energy loss rate. Dashed curves represent the theoretical expectation
for the losses in the mass loaded cases $\dot{J}\propto
\dot{M}^{1/3}$ and  $\dot{H}\propto
\dot{M}$. Continuous curves represent the best power-law fit of
the 2D monopole of Fig.~\ref{fig:5}.  The dotted lines are the
force-free solution. The squares mark cases B1 and B2, which have 
different rotation rates (see Tab.~\ref{table:1}).}
\label{fig:6}
\end{figure}

In Fig.~\ref{fig:6} the angular momentum and energy loss rates are plotted. 
The continuous lines in this figure are not a fit to our dipole results,
but are simply {\it the same curves as in Fig.~\ref{fig:5} for the 2D
monopole}. In addition, as in Fig.~\ref{fig:5}, the dashed lines are the 
analytic expectation for the loss rates in the mass-loaded limit {\it for the
monopole} (\S\ref{sec:massloaded}). 
This shows that if the solutions are parameterized in terms
of the open magnetic flux, then the dipole and monopole winds have
very similar behavior within the parameter space we have
investigated. This can be easily understood if one considers the
structure of the outflow in the far region. Given that $\dot{H}$ and
$\dot{J}$ are integrals, they can be evaluated at any distance from
the star. Even if the the field is dipolar close to the star it is
nearly monopolar outside $R_L$. Even when we do not impose our
steady-state boundary conditions at the equator, and plasmoids are
present during the evolution, $\dot{H}$ and $\dot{J}$ closely follow
our results for the monopole (again, as long as these losses are
written in terms of the open magnetic flux).  Note also that in Figure
\ref{fig:6}, we find that $\dot{H}/\Phi_o^2\Omega^2$ and
$\dot{J}/\Phi_o^2\Omega$ converge to the expected value of 2/3 in the
force-free limit (\citealt{cont99}, \citealt{gruz05}).  However,
as we discuss below, $r_Y < R_L$ in all of our calculations, contrary
to the assumption that $r_Y = R_L$ in the above force-free treatments.

We can extrapolate our results to the force-free limit for the spin-down rate.
In terms of the equivalent surface magnetic field, and using Eq.~(\ref{equiv}) with
$r_Y=R_L$ and $B_o=0.5B_r(r_{NS},\theta=0)$ we have:
\ba
\dot{H}&=&\frac{2}{3}\Omega^2r_{NS}^4B_{r-equiv}(r_{NS})^2\nonumber\\
&\approx&\frac{2}{3}\Omega^2r_{NS}4B_o^2\left(\frac{3}{5}\frac{r_{NS}}{R_L}\right)^2\nonumber\\
&\approx&\frac{24}{25}\mu^2\Omega^4,
\label{forcefreehdot}
\ea
in agreement with \citet{gruz05}.

\begin{figure*}
\resizebox{\hsize}{!}{\includegraphics[clip=true]{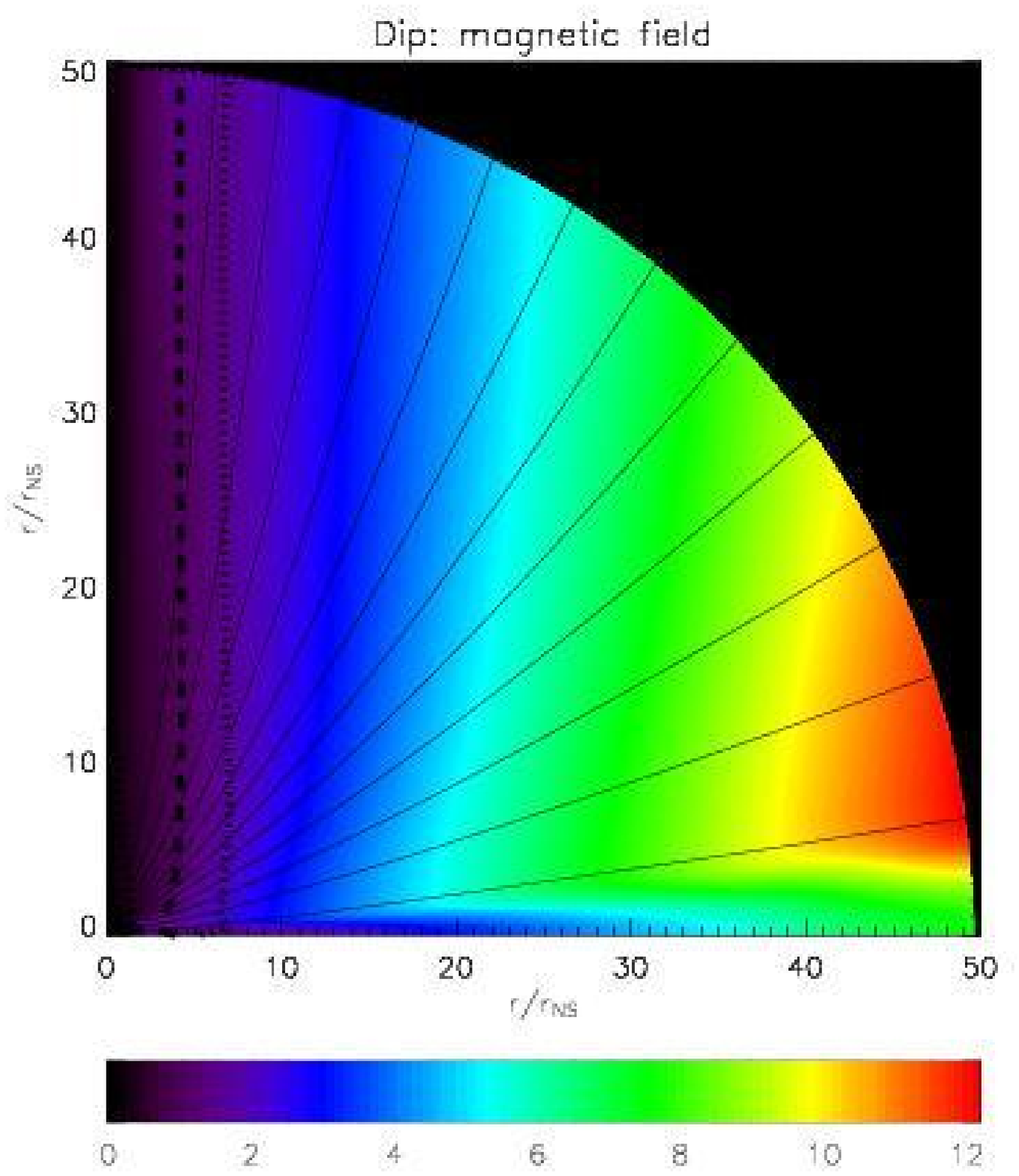}\includegraphics[clip=true]{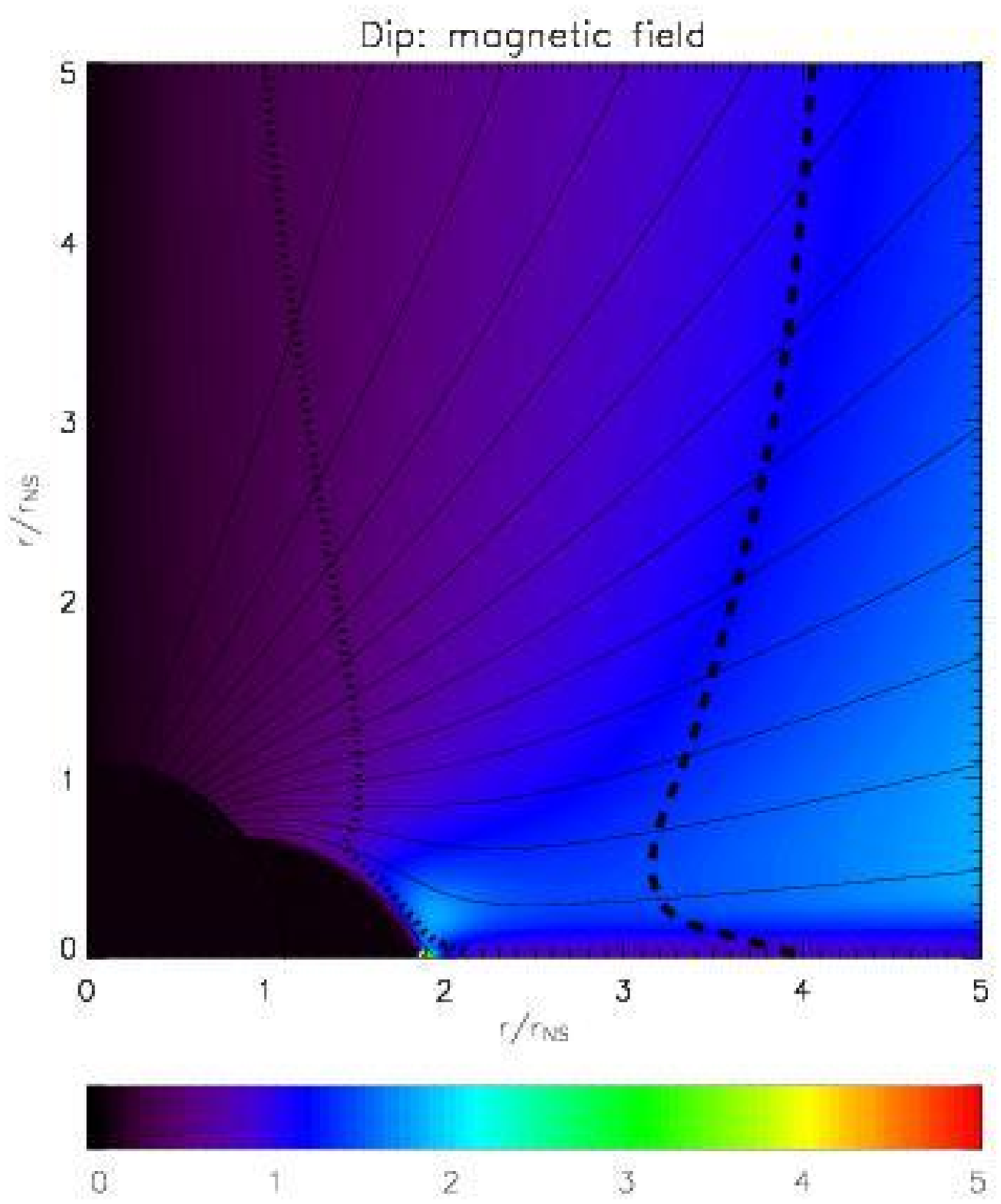}
\includegraphics[clip=true]{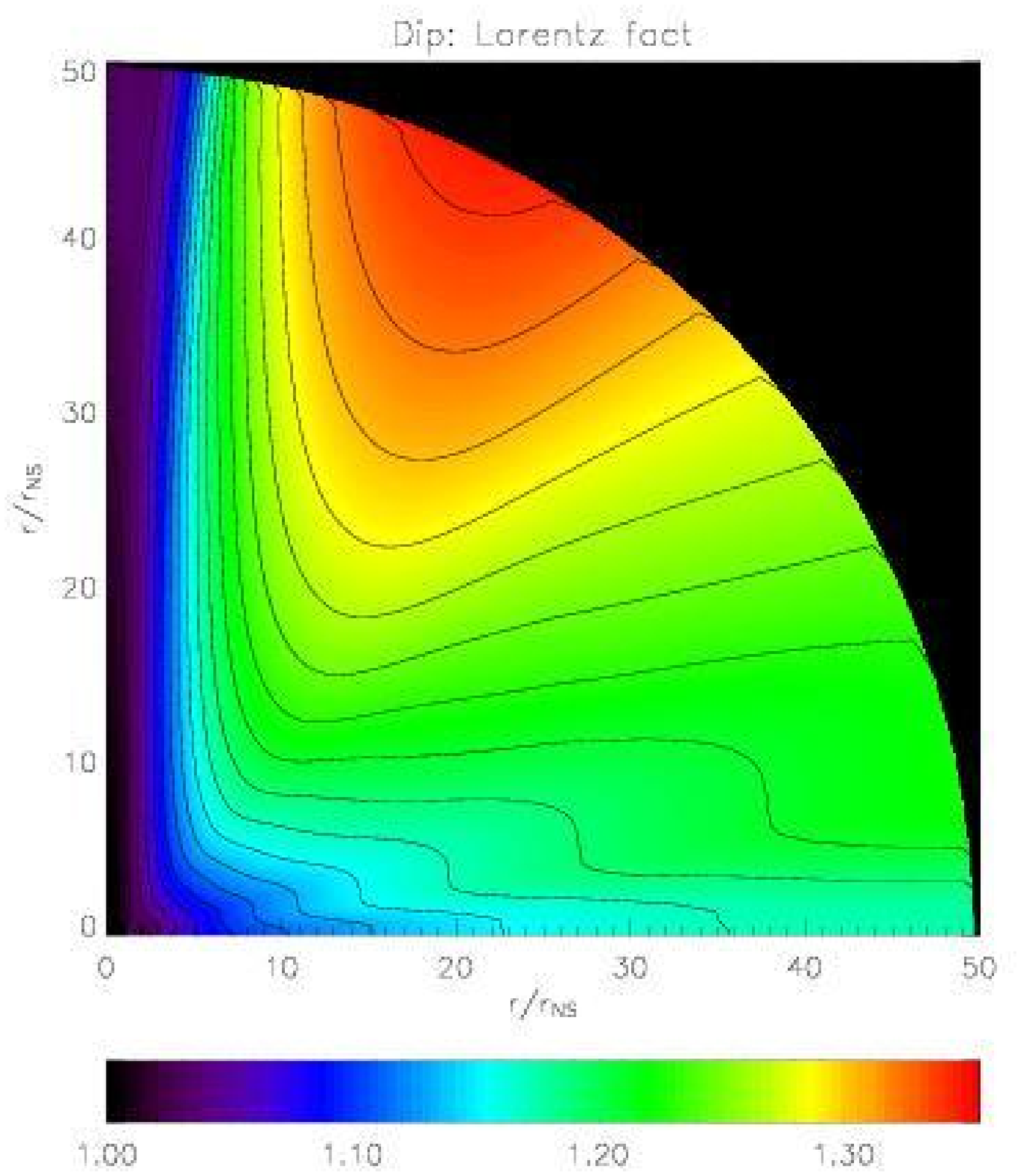}}
\resizebox{\hsize}{!}{\includegraphics[clip=true]{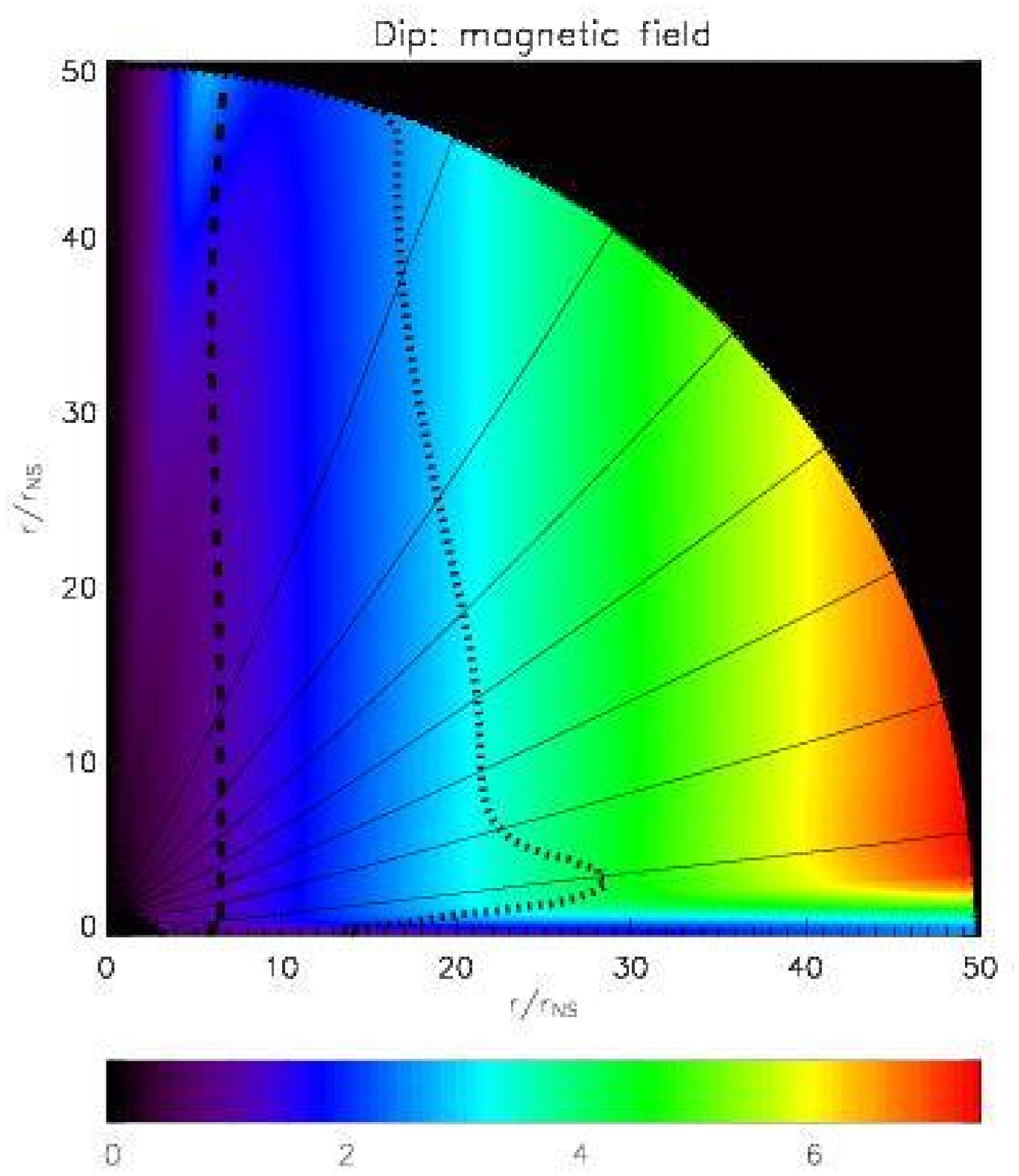}\includegraphics[clip=true]{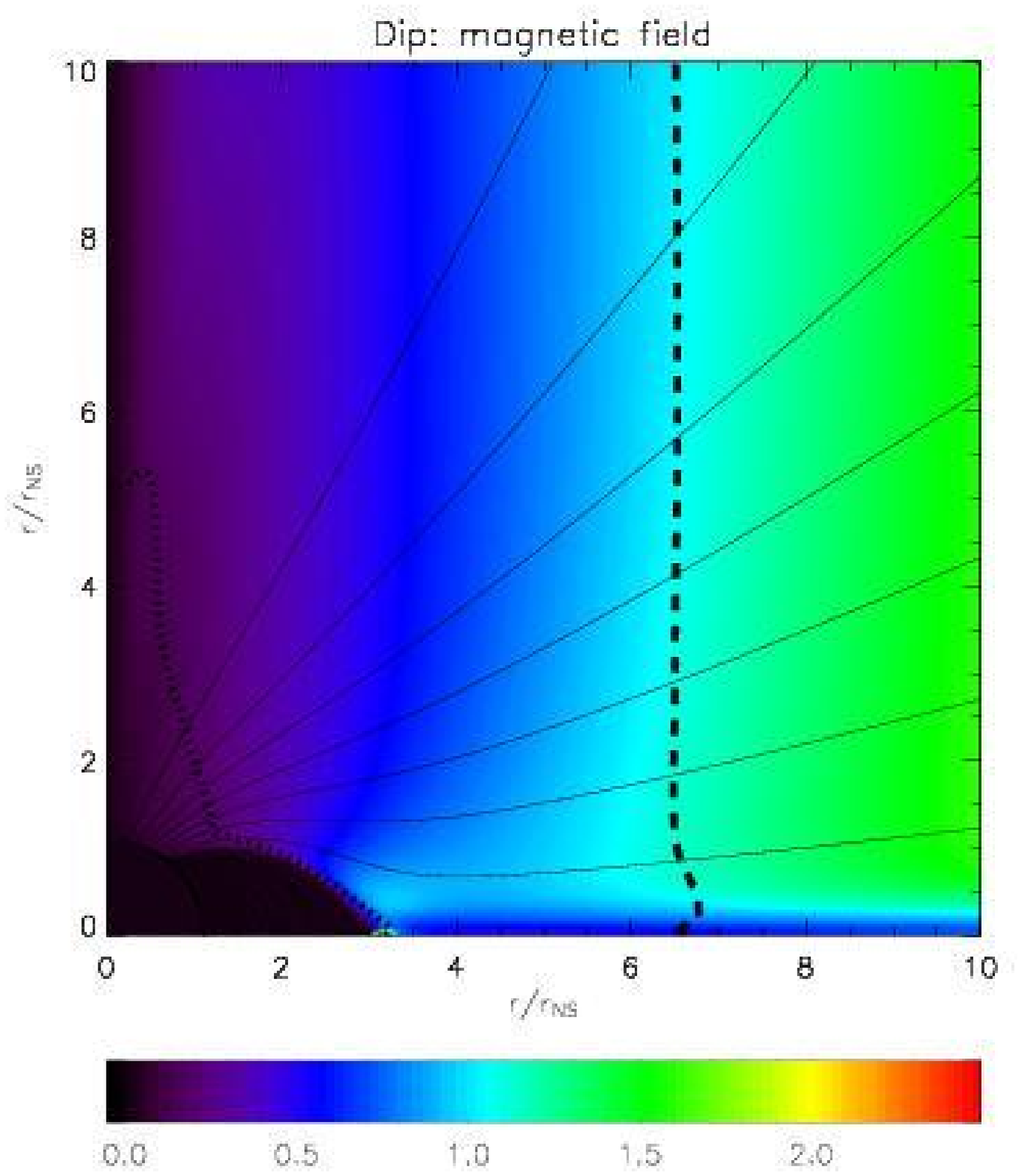}
\includegraphics[clip=true]{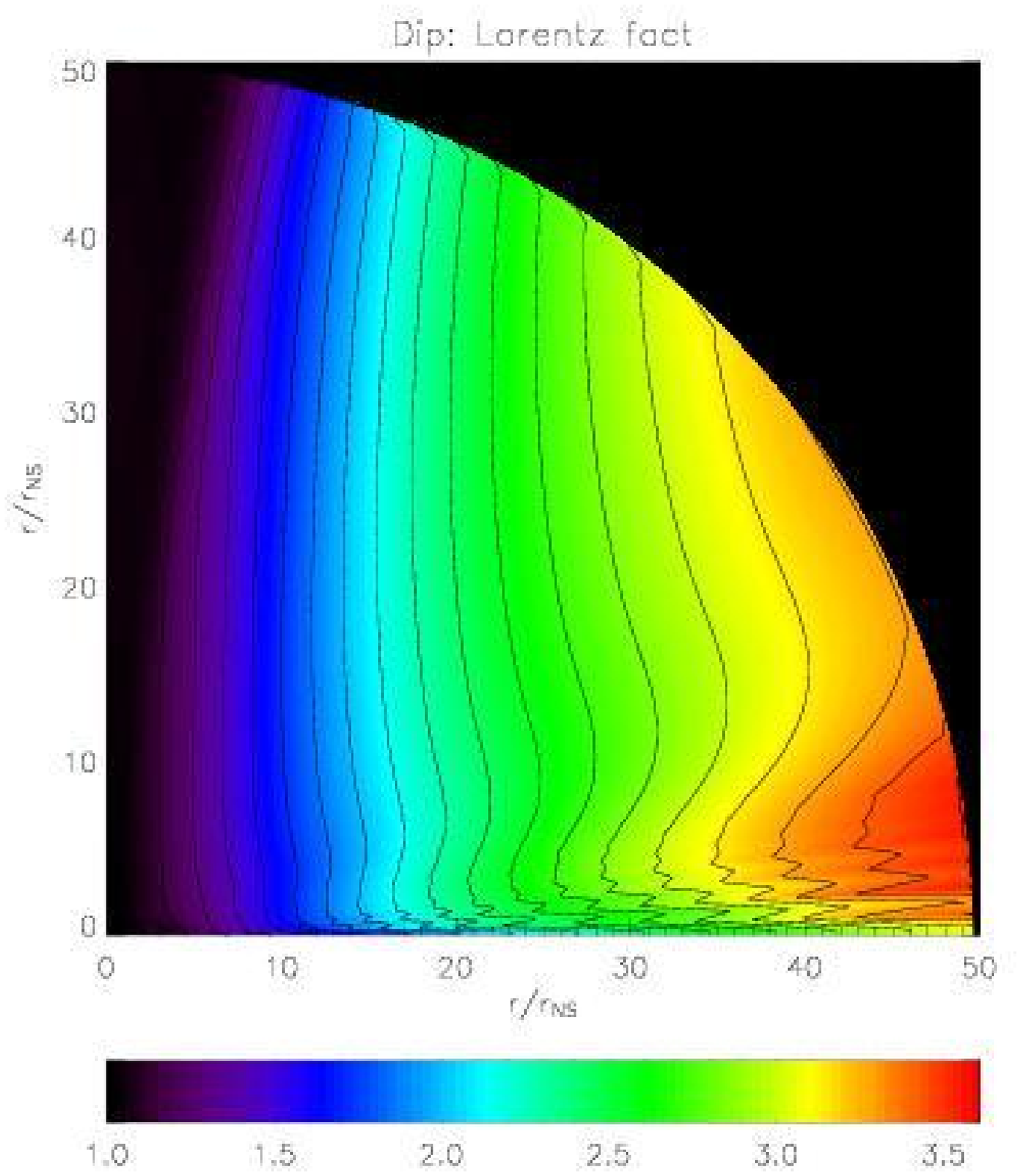}}
\caption{Result of the 2D dipole. Upper panels represent case A, lower panels 
   case C . Left: contours represents poloidal magnetic field lines, while colors 
   represent the ratio $B_{\phi}/B_r$. The dotted line is the FM  surface, the 
   dashed line the AL surface. In case C the AL surface is close to $R_{L}$. Middle: 
   contours represents poloidal magnetic field lines, while colors represent the ratio 
   $B_{\phi}/B_r$, in a region closer to the star. Now the dotted line represents the 
   SM surface. Right: colors and contour represent the Lorentz factor.}
\label{fig:7}
\end{figure*}

The flow structure for both low $\sigma_0$ (Case A) and high $\sigma_0$
(Case C) are shown in Figure \ref{fig:7}. These figures show that the
poloidal magnetic field for $R>R_L$ has a structure very similar to
that obtained from the monopole calculations. Also, $B_\phi$ scales as
$\sin{\theta}$, except in a region close to the equatorial plane where
it changes sign. This agrees with the results of Fig.~\ref{fig:6}
showing that the energy and angular momentum losses scale as for a
monopole.

As expected, we find that the size of the closed region increases with
magnetization. The position of the $Y$ point moves from $1.8 r_{NS}$
in Case A, to $2.5 r_{NS}$ in Case B, to $3.1 r_{NS}$ in Case C. Thus,
an increase in $\sigma_0$ of a factor of $\sim$60 corresponds to a
$\sim$70\% increase in the size of the closed zone. Importantly, even
at relatively high $\sigma_0$ ($\sim$ 18 for Case C), these values are
significantly smaller than the Light Cylinder radius $R_L=6.8r_{NS}$. In
order to understand the systematics of the $Y$ point, we have
calculated several models with different rotation rates $\Omega$ (see
Tabs.~\ref{table:1} \& \ref{table:4}). The radius of the $Y$ point
changes from $2.2 r_{NS}$ in Case B1, to $2.5 r_{NS}$ in Case B, to
$4.3 r_{NS}$ in Case B2.  Although $r_Y$ is largest in the model with
slowest rotation (Case B2), this model has the {\it smallest} ratio 
$r_Y/R_L$. Indeed, we find that $r_Y/R_L$ decreases as $\Omega$ decreases.
This trend yields a braking index less than 3. If one takes Cases B1, B, and B2 
as a time-series in the life of a neutron star, one would infer
a braking index $\sim$2.2. 

In general the size of the closed zone will depend on the physical
conditions at the stellar surface, including the thermal sound speed
and the mass density, which govern the mass loss rate on each open streamline
(see \S\ref{sec:poynting}, \citealt{mestel87}). This shows that even 
a small thermal pressure can have an important effect on the torque. 
Figure \ref{fig:7} shows that 
in Case A both the AL and the FM surfaces are inside
the radius of the Light Cylinder. However, in the high-$\sigma_0$ Case C
the AL surface is very close to $R_L$. This suggests that although the
AL surface rapidly approaches $R_L$ as $\sigma_0$ increases, the
position of the $Y$ point remains inside $R_L$ and is a weak function
of $\sigma_0$. It is thus possible to produce a relativistic outflow,
even if the $Y$ point is well inside $R_L$.  Because the range of
parameters we have explored is fairly limited, we can only conclude
 that quite large magnetization is required to achieve $r_Y=R_L$.
A rough extrapolation of our results implies that
$B_r(r_{in},\theta=0)^2/\rho(r_{in})$ must be of order
$10^5-10^6$ to achieve $r_Y\approx R_L$, but at the moment we deem premature to draw
strong conclusions. 

We stress here that the magnetization parameter $\sigma_0$ is defined as an
integral average.  It is known that in a dipolar field geometry with 
sub-slow magnetosonic injection, the mass flux at the edge of the 
closed zone is higher than the average integrated value over the entire 
star (\citealt{kopp76}).  Thus, the magnetization varies from high to low latitudes.  
For example, in our Case C the magnetization on the open flux tubes immediately
adjacent to the closed zone is less than 7, while the $\sigma_0$ parameter 
for this model is 17.5.  Because the position of the $Y$ point and the shape
of the magnetosphere depend on local equilibrium between the closed and open
zones, this strong variation in magnetization in latitude may help explain
why even if the global flow has $\sigma_0\gg1$, $r_Y$ is less than $R_L$.

The upper and lower far-right panels of Fig.~\ref{fig:7} show that for
the Lorentz factor we recover similar behavior as for the monopole
(compare with Figs.~\ref{fig:3} \& \ref{fig:4}). In Case A, the Lorentz
factor reaches its maximum at about $70^\circ$ from the equator. We
also notice that there is now a slower equatorial flow corresponding
to what is known as the ``slow solar wind'' in models of the Sun's
outflow. In Case C the Lorentz factor appears to scale mostly as the
cylindrical radius. This again was found for the monopole. The slow
wind region is still present, but now the boundary conditions we have
imposed on the equator to suppress the plasmoids cause a noisy
structure in $v_r$. 

The shape of the field lines outside
the FM surface does show significant differences between the
monopole and the dipole. In the dipolar Case C at $r=50r_{NS}$ the maximum
value of $B_\theta/B_r=-0.06$. Such a value for this ratio is in
between Cases C ($-0.08$) and D ($-0.03$) for the 2D monopole.  Our
limited computational domain prohibits a more quantitative study
of the poloidal field line shape at  still larger distances from the neutron star.

Lastly, the angular distribution of the energy and momentum loss rates
in our dipole models are qualitatively similar to those we obtained
for the 2D monopole (see the bottom panels of Figs.~\ref{fig:3} \&
\ref{fig:4}). For low-$\sigma_0$ outflows, because of hoop stress,
$\dot{H}$ and $\dot{J}$ far from the neutron star are peaked at high
latitudes, along the axis of rotation.  In contrast, for
high-$\sigma_0$ outflows, $\dot{H}$ and $\dot{J}$ are maximum near the
equator, similar to Fig.~\ref{fig:4} (Case E).


\section{Discussion \& Conclusions}
\label{section:discussion}

We have solved for the dynamics of time-dependent relativistic MHD
winds from rotating neutron stars, including the effects of general relativity
on the gravitational force.  The mass loss rate along open field lines is
derived self-consistently as a solution to the RMHD equations, subject to
the boundary conditions at the stellar surface (finite thermal pressure)
and a polytropic equation of state.
We consider 1D and 2D monopole field geometries and
the aligned dipole, and we explore solutions that cover the parameter
regime from non-relativistic mass-loaded low-$\sigma_0$ outflows to
relativistic Poynting flux dominated high-$\sigma_0$ outflows.  Our
primary results are:
\begin{enumerate}
\item In the 1D and 2D monopole calculations, we reproduce the
expected analytic trends in both the high-$\sigma_0$ and low-$\sigma_0$
limits.  In particular, the solutions asymptotically approach the
force-free limit when $\sigma_0 \gg1$.
\item In both the dipole and 2D monopole solutions, when $\sigma_0<1$,
the energy and momentum losses far from the neutron star
are highly directed along the axis of rotation.  The zenith angle at
which these fluxes are maximized is an increasing function of $\sigma_0$
so that for $\sigma_0>1$, the losses are primarily equatorial.
\item In both the dipole and 2D monopole solutions, for winds with
$\sigma_0<1$, the Lorentz factor peaks at high latitudes.
\item For the aligned dipole, the equatorial current
sheet may be unstable to the formation of plasmoids, leading to
time-dependent spindown of the neutron star.  A proper treatment
of dissipative RMHD in the equatorial region is needed to explore this
issue more completely.
\item If the energy and angular momentum losses from the aligned
dipole are parameterized in terms of the open magnetic flux, then the
results are nearly identical to those from the 2D monopole solutions
in both the mass-loaded and Poynting flux-dominated regimes.  In particular,
the dipole calculations quantitatively approach the force-free limit when $\sigma_0\gg1$.  
The normalization for the angular momentum loss rate is $k \approx 24/25$ 
(eq.~\ref{forcefreehdot}), in good agreement with the force-free results 
found by \citet{gruz05} and \citet{spit05}.
\item For average integrated magnetization parameters as high as 
$\sigma_0\approx20$, the radial position of the $Y$ point ($r_Y$), which 
bounds the closed zone in the dipole models, is significantly less than 
the Light Cylinder distance. This result obtains despite the fact that the 
Alfv\'en point rapidly approaches $R_L$ for $\sigma > 1$.
The ratio $r_Y/R_L$ is a very slowly increasing 
function of the surface magnetic field strength. 
An extrapolation of our results indicates that 
$\sigma_0$ must be very large, $\sigma_0 \gg 100$, for the $Y$ point to
reach the Light Cylinder.   
That $r_Y$ is generically much less than $R_L$ in our calculations is due
in part to the fact that although $\sigma_0\approx20$ globally 
(Case C, \S\ref{sec:dipole}) the magnetization along the open field lines
nearest the closed zone is less than 7. \label{item:ryrl}
\item Over limited dynamic range, the ratio
$r_Y/R_L$ decreases as $\omega$ decreases in the dipole models. This 
behavior is consistent with a braking index less than 3 
(see \S\ref{sec:poynting} and  \S\ref{section:pulsars}). \label{item:omega}
\end{enumerate}
Concerning the acceleration of the wind, we do not find any evidence for efficient
magnetic to kinetic energy conversion. However we want to stress again that our 
computational box does not extend far enough from the star to appreciably probe the 
asymptotic behavior of the wind. 


\subsection{Rotation-Powered Pulsars}
\label{section:pulsars}

Our results have several possible implications for rotation powered
pulsars, where $\sigma_0$ is high. The fact that the Y-point is
 interior to the light cylinder (and the
Alfven radius) suggests that observed braking indices less than 3
might be a consequence of equilibrium magnetospheric structure.
We emphasize that the position of the $Y$ point depends on a local equilibrium
between the surface of the closed zone and the wind region.  As a consequence,
the value of the mass flux along the open field lines closest to the closed 
zone directly affects the position of the $Y$ point.  The mass flux, in turn,
depends on the boundary conditions at the stellar surface.  In our simulations,
the mass loss rate is derived self-consistently, subject to a finite thermal
pressure at $r_{NS}$.  In the
case of a pulsar,  the injection of matter at $r_{NS}$ is 
thought to be due to non-MHD pair creation processes (\citealt{hibsch01})
in a ``gap'' just above the stellar surface which,
in an otherwise MHD flow, act to inject plasma with velocity already exceeding
the sound speed. Then  
the mass loss rate is determined by the gap physics, not by the requirement of
making a smooth transition from subsonic to supersonic flow.
 Thus, in principle $r_Y$ depends upon the injection law determined by the
 pair creation physics at the surface. For this reason,  we 
caution against over-interpreting our results \ref{item:ryrl} and \ref{item:omega} 
in the pulsar context.  Indeed, thermal pressure of the magnitude employed here is not 
likely to be relevant for classical pulsars.\footnote{If the surface magnetic field is 
not highly stressed, the plasma in the closed zone is likely to be
non-neutral and it will fill the magnetosphere via cross-field transport driven by
shear flow turbulence (\citealt{spit02}), not via pressure, which alters
the force balance at the $Y$ point. }  
On the other hand, magnetic dissipation and reconnection at and 
near the $Y$ point might cause pressure and inertial forces to be significant in this localized
region. In any case, our primary conclusions, that $r_Y/R_L$ is 
generally less than unity and that  $r_Y/R_L$  decreases as $\Omega$ decreases, 
are intriguing possibilities now open to investigation with the advent of dynamical,
large $\sigma$ models of neutron star magnetopsheres.
A more general study of the effect of the injection conditions on the structure 
of the magnetosphere and the accompanying wind is under way.\footnote{
We are aware of the recent work of Komissarov (2005) who obtains $r_Y\approx R_L$
for $\sigma\approx100$.  Because the injection conditions used by Komissarov (2005)
are qualitatively different from the self-consistent calculation of the mass loss
rate as a function of latitude obtained here, we do not believe our two results are
mutually contradictory.  Instead, our finding that $r_Y\approx R_L$ only when 
$\sigma\gg100$ serves to emphasize the point that the injection conditions 
are critical in determining $r_Y$.}

The appearance of outwardly propagating plasmoids at and beyond the
Y-point as a consequence of (numerical) magnetic dissipation raises
the intriguing possibility that noise in pulsar spindown
\citep{cordes80} might arise from instability of the magnetospheric
currents due to real magnetic dissipation (e.g., via the collisionless
tearing instability).  Our results show the possibility of 20\% or
more torque fluctuations that could in principle give rise to a random
walk in the rotation frequencies of pulsars, as is observed.
Likewise, such torque fluctuations might give rise to noise and limit
cycles in the observed phase of pulsars' subpulses, as is seen in many
systems (\citealt{rankin86}, \citealt{desh99}).  Determining whether
these observed phenomena could be due to magnetospheric dissipation
requires treating the dissipation with a consistent physical model,
which is an investigation beyond that reported in this paper.


\subsection{Proto-Neutron Stars \& Proto-Magnetars}

High thermal pressure at the neutron star surface is thought to be a generic feature
of neutron star birth and so the calculations presented in this paper 
are directly relevant to young neutron stars, particularly in the hypothesized 
rapidly-rotating and highly-magnetic initial state of magnetars.

We identify five separate phases in any very young 
neutron star's life: (1) a pressure-dominated essentially non-magnetic phase in
which the wind is driven by neutrino-heating (as in e.g., \citealt{qian96}), 
(2) a phase in which magnetic field effects are present,
but not dominant so that $r_A\omega<0.1c\approx c_T$,
where $c_T$ is the isothermal sound speed at the proto-neutron star surface,
(3) a non-relativistic magnetically-dominated phase when 
$R_A$ is greater than $r_{NS}$, but less than $R_L$, (4)
a relativistic phase in which $R_A\sim R_L$, but $r_Y<R_L$,
and lastly (5) an epoch when the force-free limit is applicable and 
$r_Y\simeq R_A\simeq R_L$.  Phases (1)$-$(5) represent a time evolution starting 
immediately after the supernova explosion commences.  The timescale for the 
evolution from phase (1) to phase (4) is set by the Kelvin-Helmholtz timescale
for cooling of the proto-neutron star, $\tau_{\rm KH}\sim10-100$ seconds.  The transition
from phase (4) to phase (5) may occur on a longer timescale, or not at all,
depending upon the applicability of our results to classical pulsars as the
MHD approximation breaks down.
 
For parameters appropriate to a proto-magnetar, phase (3) lasts of order $\tau_{\rm KH}$.
Our simulations show that in this phase, the wind is energetic and that the energetic flux 
is highly directed along the axis of rotation.  The characteristic rotational energy
loss rate in this phase is 
$\dot{E}\approx 4\times10^{49}\, 
B_{14}^2 P_{1}^{-5/3} \,\,\,{\rm ergs\,\,s^{-1}}$,
where $B_{14}=B(r_{NS})/10^{14}$ G is the ``equivalent'' 
monopole field (eq.~\ref{equiv}), and $P_{1}=P/1$ ms.
On the timescale $\tau_{\rm KH}$, the total amount of energy 
extracted is comparable to the asymptotic supernova energy, $\sim10^{51}$ ergs. 
The magnitude of the rotational energy extracted in this phase and its collimation along
the rotational axis should have profound implications both for the spindown of 
millisecond magnetars and for the supernova remnants that accompany their birth.
We save a detailed discussion for a future paper.


\section*{acknowledgements}

We thank Prof. S.~S. Komissarov, for providing us with a wave based Riemann solver, to verify effects of numerical diffusion in our 2D dipole calculations. Part of this work was done while N.~B. and J.~A. were in residence at the Kavli Institute 
for Theoretical Physics, with partial support from NSF Grant No.PHY99-07949. 
N.~B. and J.~A. were also supported in part by NASA grants TM4-5000X and NAG5-12031, and
by NSF grant AST-0507813 to the University of California, Berkeley.  E.~Q. was supported in part by the  David and Lucile Packard Foundation. T.A.T. was supported during most of this work by NASA through Hubble Fellowship grant \#HST-HF-01157.01-A awarded by the Space Telescope Science Institute, operated by the Association of Universities for Research in Astronomy, Inc., for NASA, under contract NAS 5-26555. 


\appendix
\section{Parametrization of the models}

All the simulations can be parametrized in terms of the ratios of characteristic velocities
at the injection radius ($r_{in}$). Following the work by \citet{mestel68a} and 
\citet{gj70}, these are esentially the sound speed $c_{T}$, the rotational
velocity $v_\phi$, the non relativistic AL velocity $\sqrt{B^2/8\pi\rho}$, 
and the escape speed $\sqrt{2GM/r}$. For the 2D cases we consider the value of 
the rotational velocity and the magnetic field at the equator. Note that the characterisitc 
velocities defined above do not coincide with the cited works in newtonian gravity. 
For example the escape 
speed differs by a factor $\sqrt{2}$, and in the definition of $v_\phi$ one must 
include the effect of time delay. These are corections due to the general relativisitc 
metric which are not present in the above cited papers ($\alpha(r_{in})=0.79$).  

Notice that for the dipole, where the open field lines originate close to the pole, the base
value of the magnetic field is about twice stronger than at the equator, and the rotational velocity
is much smaller than the equatorial speed.

\begin{table*}
\begin{minipage}{17cm}
\caption{Parametrization of the numerical models.}
\label{table:5}
\begin{center}
\begin{tabular}{c c c c c c c c c c c}
\hline
1D Mod. & A0 & A & B & C & D & E & F & G & H & I\\
\hline
$\frac{2GM}{r_{in}c^2_T}$ & 11.44 & 11.44 & 6.74 & 7.54 & 8.98 & 10.45 & 11.44 & 6.29 & 6.29 & 6.29\\
$\frac{B_r^2}{8\pi\rho c^2_T}$ & 0.061 & 0.61 & 0.71 & 0.80 & 0.95 & 1.11 & 60.6 & 133 & 533 & 1330 \\
$\frac{2GM}{v_\phi^2r_{in}}$ & 9.43 & 9.43 & 9.43 & 9.43 & 9.43 & 9.43 & 9.43 & 9.43 & 9.43 & 9.43 \\
\hline
2D Monop. Mod. & A & B & B1 & C & D & E & & &  \\
\hline
$\frac{2GM}{r_{in}c^2_T}$ & 11.44 & 11.44 & 11.44 & 11.44 & 11.44 & 11.44 &  &  &  \\
$\frac{B_r^2}{8\pi\rho c^2_T}$ & 0.61 & 6.1 & 6.1 & 60.6 & 606 & 3030 &  &  &  \\
$\frac{2GM}{v_\phi^2r_{in}}$ & 9.43 & 9.43 & 4.20 & 9.43 & 9.43 & 9.43 &  &  & \\
\hline
2D Dip. Mod. & A & B & B1 & B2 & C &  &  & & \\
\hline
$\frac{2GM}{r_{in}c^2_T}$ & 11.44 & 11.44 & 11.44 & 11.44 & 11.44 &  &  &  &  \\
$\frac{B_\theta^2}{8\pi\rho c^2_T}$ & 1.22 & 12.2 & 12.2 & 12.2 & 122 &  &  &  &  \\
$\frac{2GM}{v_\phi^2r_{in}}$ & 9.43 & 9.43 & 4.20 & 38.5 & 9.43 &  &  &  & \\
\hline
\end{tabular}
\end{center}
\medskip

\end{minipage}
\end{table*}

\bsp

\label{lastpage}


\begin{thebibliography}{}


\bibitem[\protect\citeauthoryear{Anile}{1989}]{anile89}
Anile, M.\ 1989, {\it Relativistic Fluids and Magneto-Fluids}, Cambridge University Press, Cambridge

\bibitem[\protect\citeauthoryear{Arons}{1981}]{arons81} 
Arons, J.\ 1981, ``The Slot Gap Model of Pulsars'', in IAU Symp. No. 95, 'Pulsars', W.~Sieber and R.~Wielebinski, eds. (Dordrecht; D.~Reidel), 69-85

\bibitem[\protect\citeauthoryear{Arons}{1983}]{arons83} 
Arons, J.\ 1983, ApJ, 266, 215

\bibitem[\protect\citeauthoryear{Arons}{2003}]{arons03} 
Arons, J.\ 2003, ApJ, 589, 871 

\bibitem[\protect\citeauthoryear{Arons}{2004}]{arons04} 
Arons, J.\ 2004, IAU Symposium, 218, 163 

\bibitem[\protect\citeauthoryear{Aschwanden et al.}{2001}]{asch01} 
Aschwanden, M.~J., Poland, A.~J., \& Rabin, D.~M.\ 2001, ARA\&A, 39, 175

\bibitem[\protect\citeauthoryear{Begelman \& Li}{1994}]{beg94} 
Begelman, M.~C., \& Li, Z.-Y.\ 1994, ApJ, 426, 269

\bibitem[\protect\citeauthoryear{Belcher \& MacGregor}{1976}]{belcher76} 
Belcher, J.~W., \& MacGregor, K.~B.\ 1976, ApJ, 210, 498 

\bibitem[\protect\citeauthoryear{Beskin et al.}{1998}]{beskin98} 
Beskin, V.~S., Kuznetsova, I.~V., \& Rafikov, R.~R.\ 1998, MNRAS, 299, 341 

\bibitem[\protect\citeauthoryear{Blandford \& Romani}{1988}]{bland88} 
Blandford, R.~D., \& Romani, R.~W.\ 1988, MNRAS, 234, 57

\bibitem[\protect\citeauthoryear{Blasi et al.}{2000}]{blasi00} 
Blasi, P., Epstein, R.~I., \& Olinto, A.~V.\ 2000, ApJL, 533, L123 

\bibitem[\protect\citeauthoryear{Bogovalov}{2001}]{bog01}
Bogovalov, S.~V.\ 2001, A\&A, 371, 1155

\bibitem[\protect\citeauthoryear{Burrows \& Lattimer}{1986}]{burrows86} 
Burrows, A., \& Lattimer, J.~M.\ 1986, ApJ, 307, 178

\bibitem[\protect\citeauthoryear{Camenzind}{1986a}]{cam86a}
Camenzind, M.\ 1986a, A\&A, 156, 137

\bibitem[\protect\citeauthoryear{Camenzind}{1986b}]{cam86b}
Camenzind, M.\ 1986b, A\&A, 162, 32

\bibitem[\protect\citeauthoryear{Camenzind}{1987}]{cam87}
Camenzind, M.\ 1987, A\&A, 184, 341

\bibitem[\protect\citeauthoryear{Camilo et al.}{2000}]{camilo00} 
Camilo, F., et al.\ 2000, ApJ, 541, 367

\bibitem[\protect\citeauthoryear{Cheng}{1987a}]{cheng87a} 
Cheng, K.~S.\ 1987a, ApJ, 321, 805 

\bibitem[\protect\citeauthoryear{Cheng}{1987b}]{cheng87b} 
Cheng, K.~S.\ 1987b, ApJ, 321, 799 

\bibitem[\protect\citeauthoryear{Contopoulos et al.}{1999}]{cont99}
Contopoulos, I., Kazanas, D., Fendt, C.\ 1999, ApJ, 511, 351

\bibitem[\protect\citeauthoryear{Cordes \& Helfand}{1980}]{cordes80} 
Cordes, J.~M., \& Helfand, D.~J.\ 1980, ApJ, 239, 640 

\bibitem[\protect\citeauthoryear{Daigne \& Drenkhahn}{2002}]{daigne02}
Daigne, F., \& Drenkhahn, G.\ 2002, A\&A, 381, 1066 

\bibitem[\protect\citeauthoryear{Deeter et al.}{1999}]{deeter99} 
Deeter, J.~E., Nagase, F., \& Boynton, P.~E.\ 1999, ApJ, 512, 300 

\bibitem[\protect\citeauthoryear{Del Zanna et al.}{2003}]{ldz03}
Del Zanna, L., Bucciantini, N., Londrillo, P.\ 2003, A\&A, 400, 397

\bibitem[\protect\citeauthoryear{Deshpandhe \& Rankin}{1999}]{desh99}
Deshpande, A.~A., \& Rankin, J.~M.\ 1999, ApJ, 524, 1008

\bibitem[\protect\citeauthoryear{Duncan et al.}{1986}]{ducan86} 
Duncan, R.~C., Shapiro, S.~L., \& Wasserman, I.\ 1986, ApJ, 309, 141 

\bibitem[\protect\citeauthoryear{Duncan \& Thompson}{1992}]{duc92} 
Duncan, R.~C., \& Thompson, C.\ 1992, ApJL 392, L9 

\bibitem[\protect\citeauthoryear{Endeve \& Leer}{2003}]{endeve03}
Endeve, E., \& Leer, E.\ 2003, ApJ, 589, 1040

\bibitem[\protect\citeauthoryear{Goldreich \& Julian}{1969}]{gj69} 
Goldreich, P., \& Julian, W.~H.\ 1969, ApJ, 157, 869  

\bibitem[\protect\citeauthoryear{Gold\-reich \& Julian}{1970}]{gj70} 
Goldreich, P., \& Julian, W.~H.\ 1970, ApJ, 160, 971  

\bibitem[\protect\citeauthoryear{Goodwin et al.}{2004}]{good04} 
Goodwin, S.~P., Mestel, J., Mestel, L., \& Wright, G.~A.~E.\ 2004, 
MNRAS, 349, 213 

\bibitem[\protect\citeauthoryear{Gosling}{1996}]{gosling96}
Gosling, J.~T.\ 1996, ARA\&A, 34, 35

\bibitem[\protect\citeauthoryear{Gruzinov}{2005}]{gruz05}
Gruzinov, A.\ 2005, Phys. Rev. Lett., 94

\bibitem[\protect\citeauthoryear{Heyvaerts \& Norman}{2003}]{hey03} 
Heyvaerts, J., \& Norman, C.\ 2003, ApJ, 596, 1240 

\bibitem[\protect\citeauthoryear{Hibschman \& Arons}{2001}]{hibsch01} 
Hibschman, J., \& Arons, J.\ 2001, ApJ, 560, 871 

\bibitem[\protect\citeauthoryear{Jin et al.}{2001}]{jin01}
Jin, S.-P., Hu, X.-P., Zong, Q.-G., et al.\ 2001, JGR, 106, 29,807

\bibitem[\protect\citeauthoryear{Kaspi et al.}{1994}]{kaspi94} 
Kaspi, V.~M., Manchester, R.~N., Siegman, B., Johnston, S., \& Lyne, A.~G.\ 1994, ApJL, 422, L83

\bibitem[\protect\citeauthoryear{Keppens \& Goedbloed}{2000}]{kep00}
Keppens, R., \& Goedbloed, J.~P.\ 2000, ApJ, 530, 1036

\bibitem[\protect\citeauthoryear{Koide et al.}{1999}]{koide99}
Koide, S., Shibata, K., Kudoh, T.\ 1999, ApJ, 522, 727

\bibitem[\protect\citeauthoryear{Komissarov}{1999}]{komissarov99}
Komissarov, S.~S.\ 1999, MNRAS, 303, 343

\bibitem[\protect\citeauthoryear{Kopp \& Holzer}{1976}]{kopp76}
Kopp, R.~A., \& Holzer, T.~E.\ 1976, Solar Physics, 49, 43

\bibitem[\protect\citeauthoryear{Landau \& Lifshitz}{1971}]{landau71}
Landau, L.~D., Lifshitz, E.~M.\ 1971, The Classical Theory of Fields, Pergamon Press, Oxford 

\bibitem[\protect\citeauthoryear{Livingstone et al.}{2005}]{livi05} 
Livingstone, M.~A., Kaspi, V.~M., Gavriil, F.~P., \& Manchester, R.~N.\ 2005, ApJ, 619, 1046 
 
\bibitem[\protect\citeauthoryear{Lyne et al.}{1993}]{lyne93} 
Lyne, A.~G., Pritchard, R.~S., \& Graham-Smith, F.\ 1993, MNRAS, 265, 1003 

\bibitem[\protect\citeauthoryear{Lyubarsky \& Eichler}{2001}]{le01} 
Lyubarsky, Y., \& Eichler, D.\ 2001, ApJ, 562, 494 

\bibitem[\protect\citeauthoryear{Mestel}{1968a}]{mestel68a} 
Mestel, L.\ 1968a, MNRAS, 138, 359 

\bibitem[\protect\citeauthoryear{Mestel}{1968b}]{mestel68b} 
Mestel, L.\ 1968b, MNRAS, 140, 177 

\bibitem[\protect\citeauthoryear{Mestel \& Spruit}{1987}]{mestel87}
Mestel, L.~\& Spruit, H.~C.\ 1987, MNRAS, 226, 57

\bibitem[\protect\citeauthoryear{Michel}{1969}]{michel69}
Michel, F.~C.\ 1969, ApJ, 158, 727

\bibitem[\protect\citeauthoryear{Michel}{1973}]{michel73}
Michel, F.~C.\ 1973, ApJ, 180, L133

\bibitem[\protect\citeauthoryear{Michel}{1991}]{michel91}
Michel, F.~C.\ 1991, Theory of Neutron Star Magnetospheres, Univ. Chicago Press, Chicago 

\bibitem[\protect\citeauthoryear{Misner et al.}{1973}]{misner73}
Misner, C.~W., Thorne, K.~S., Wheeler, J.~A.\ 1973, Gravitation, W.~H. Freeman and Company, San Francisco

\bibitem[\protect\citeauthoryear{Muslimov \& Tsygan}{1992}]{muslimov92}
Muslimov, A.~G., \& Tsygan, A.~I.\ 1992, MNRAS, 255, 61

\bibitem[\protect\citeauthoryear{Parker}{1958}]{parker58}
Parker, E.~N.\ 1958, ApJ, 128, 664

\bibitem[\protect\citeauthoryear{Pons et al.}{1999}]{pons99} 
Pons, J.~A., Reddy, S., Prakash, M., Lattimer, J.~M., \& Miralles, J.~A.\ 1999, ApJ, 513, 780 

\bibitem[\protect\citeauthoryear{Qian \& Woosley}{1996}]{qian96} 
Qian, Y.-Z., \& Woosley, S.~E.\ 1996, ApJ, 471, 331 

\bibitem[\protect\citeauthoryear{Rankin}{1986}]{rankin86}
Rankin, J. \ 1986, ApJ, 301, 901

\bibitem[Sakurai(1985)]{sak85} 
Sakurai, T.\ 1985, A\&A, 152, 121 

\bibitem[\protect\citeauthoryear{Scharlemann \& Wagoner}{1973}]{sch73} 
Scharlemann, E.~T., \& Wagoner, R.~V.\ 1973, ApJ, 182, 951 

\bibitem[\protect\citeauthoryear{Schatzman}{1962}]{sch62} 
Schatzman, E.\ 1962, Annales d'Astrophysique, 25, 18 

\bibitem[\protect\citeauthoryear{Smith}{1998}]{smi98} 
Smith, M.~D.\ 1998, Ap\&SS, 261, 169

\bibitem[\protect\citeauthoryear{Spitkovsky \& Arons}{2002}]{spit02} 
Spitkovsky, A., and Arons, J.\ 2002, in ``Neutron Stars in Supernova Remnants'', 
  P.O. Slane and B.M. Gaensler, eds. ASP Conference Series, Vol. 271
  (San Francisco: ASP), 81

\bibitem[\protect\citeauthoryear{Spitkovsky \& Arons}{2004}]{spit04} 
Spitkovsky, A., and Arons, J.\ 2004, ApJ, 603, 669

\bibitem[\protect\citeauthoryear{Spitkovsky}{2005}]{spit05}
Spitkovsky, A., in preparation

\bibitem[\protect\citeauthoryear{Tanuma \& Shibata}{2005}]{tanuma05}
Tanuma, S., \& Shibata, K.\ 2005, ApJ, 628, L77

\bibitem[\protect\citeauthoryear{Thompson \& Duncan}{1993}]{thom93} 
Thompson, C., \& Duncan, R.~C.\ 1993, ApJ, 408, 194 

\bibitem[\protect\citeauthoryear{Thompson}{1994}]{thom94} 
Thompson, C.\ 1994, MNRAS, 270, 480 

\bibitem[\protect\citeauthoryear{Thompson et al.}{2001}]{thom01}
Thompson, T.~A., Burrows, A., Meyer, B.~S.\ 2001, ApJ, 562, 887

\bibitem[\protect\citeauthoryear{Thompson}{2003}]{thom03} 
Thompson, T.~A.\ 2003, ApJL, 585, L33

\bibitem[\protect\citeauthoryear{Thompson et al.}{2004}]{thom04}
Thompson, T.~A., Chang, P., Quataert, E.\ 2004, ApJ, 611, 380

\bibitem[\protect\citeauthoryear{Timokhin}{2005}]{timo05}
Timokhin, A.\ 2005, proceedings of 
"Astrophysical Sources of High Energy Particles and Radiation", Torun, Poland, Astro-Ph/0507054.

\bibitem[\protect\citeauthoryear{Usov}{1992}]{usov92} 
Usov, V.~V.\ 1992, Nature, 357, 472 

\bibitem[\protect\citeauthoryear{Vlahakis \& K\"onigl}{2003}]{vl03} 
Vlahakis, N.~\& K\"onigl, A.\ 2003, ApJ, 596, 1080 

\bibitem[\protect\citeauthoryear{Vlahakis}{2004}]{vl04} 
Vlahakis, N.\ 2004, ApJ, 600, 324

\bibitem[\protect\citeauthoryear{Wasserman \& Shapiro}{1983}]{wasser83}
Wasserman, I., \& Shapiro, S.~L.\ 1983, ApJ, 265, 1036

\bibitem[\protect\citeauthoryear{Weber \& Davis}{1967}]{weber67}
Weber, E.~J., \& Davis, L.~Jr.\ 1976, ApJ, 148, 217

\bibitem[\protect\citeauthoryear{Weinberg}{1972}]{weinberg72}
Weinberg, S.\ 1972, Gravitation and Cosmology, J~Wiley \& Sons, New York

\bibitem[\protect\citeauthoryear{Wheeler et al.}{2000}]{wheeler00} 
Wheeler, J.~C., Yi, I., H{\"o}flich, P., \& Wang, L.\ 2000, ApJ, 537, 810 

\bibitem[\protect\citeauthoryear{Wright}{2001}]{wright01} 
Wright, G.A.E.\ 2001, MNRAS, 344, 1041

\bibitem[\protect\citeauthoryear{Yin et al.}{2000}]{yin00}
Yin, L., Coroniti, F.~V., Pritchett, P.~L., Frank, L.~A., Paterson, W.~R.\ 2000, JGR, 105, 25,345


\end{thebibliography}
\end{document}